\documentclass[preprint,10pt,authoryear]{elsarticle}

\usepackage{cite}
\usepackage{amsmath,amssymb,amsfonts}
\usepackage[ruled]{algorithm2e}
\usepackage{graphicx}
\usepackage{textcomp}
\usepackage{xcolor}
\usepackage[breaklinks, hidelinks]{hyperref}
\usepackage{tabularx}
\usepackage{comment}
\usepackage{listings}
\usepackage{multirow}
\usepackage[T1]{fontenc}
\usepackage{algpseudocode}

\SetAlCapFnt{\footnotesize}
\SetAlCapNameFnt{\footnotesize}
\SetAlCapSty{}

\lstdefinelanguage{JavaScript}{
  keywords={typeof, new, true, false, catch, function, return, null, catch, switch, var, if, in, while, do, else, case, break},
  keywordstyle=\color{blue}\bfseries,
  ndkeywords={class, export, boolean, throw, implements, import, this},
  ndkeywordstyle=\color{darkgray}\bfseries,
  identifierstyle=\color{black},
  sensitive=false,
  comment=[l]{//},
  morecomment=[s]{/*}{*/},
  commentstyle=\color{purple}\ttfamily,
  stringstyle=\color{red}\ttfamily,
  morestring=[b]',
  morestring=[b]"
}

\def\BibTeX{{\rm B\kern-.05em{\sc i\kern-.025em b}\kern-.08em
    T\kern-.1667em\lower.7ex\hbox{E}\kern-.125emX}}

\journal{Journal of Systems and Software}
\begin{document}
\begin{frontmatter}

\title{Tracking Feature Usage: Incorporating Implicit Feedback into Software Product Lines }

\author[label1]{Oscar Díaz}
\ead{oscar.diaz@ehu.eus}
\author[label1]{Raul Medeiros\corref{cor1}}
\ead{raul.medeiros@ehu.eus}
\cortext[cor1]{Corresponding author}
\affiliation[label1]{organization={University of the Basque Country (UPV/EHU)},
            city={San Sebastián},
            state={Gipuzkoa},
            country={Spain}}
\author[label2]{Mustafa Al-Hajjaji}
\ead{mustafa.alhajjaji@pure-systems.com}
\affiliation[label2]{organization={pure-systems GmbH},
            city={Magdeburg},
            country={Germany}}

\begin{abstract}
Implicit feedback is  collecting information about software usage to understand how and when the software is used. This research tackles implicit feedback in Software Product Lines (SPLs).  The need for platform-centric feedback makes SPL feedback depart from one-off-application feedback in both the artefact to be tracked (the platform vs the variant) as well as the tracking approach (indirect coding vs direct coding). Traditionally, product feedback is achieved by embedding `usage trackers'  into the software's code. Yet, products are now members of the SPL portfolio, and hence, this approach conflicts with one of the main SPL tenants:  reducing, if not eliminating, coding directly into the variant's code. 
Thus, we advocate for  Product Derivation  to be subject to a second transformation that precedes the construction of the variant based on the configuration model.   This approach is tested through \textit{FEACKER},  an extension to \textit{pure::variants}. We resorted to a TAM  evaluation on  \textit{pure-systems GmbH}  employees(n=8). Observed divergences were next tackled through a focus group  (n=3). The results reveal agreement in the interest in conducting feedback analysis at the platform level (perceived usefulness) while regarding FEACKER as a seamless extension to \textit{pure::variants}' gestures (perceived ease of use).

\end{abstract}

\end{frontmatter}

\section{Introduction}

Continuous deployment is considered an extension of Agile practices whereby  the user feedback  is extended beyond development by frequently releasing and deploying software to customers and continuously learning from their usage \citep{Dakkak2021a}. 
Feedback can be obtained either directly, e.g., through reviews (i.e., explicit feedback) or through tracking user interactions (i.e., implicit feedback). We focus on implicit feedback, i.e., the automatically collected information about software usage from which user behavior and preferences can be inferred \citep{Oordt2021}. 

In one-off development, implicit feedback (hereafter just `feedback') benefits developers and analysts alike. Developers can benefit from a more timely awareness of bugs, bad smells, or usability issues \citep{Johanssen2019} while being motivated by the fact that their software is bringing real value to real users \citep{Oordt2021}. Regarding the analysts, they resort to  feedback for requirement prioritization \citep{Wang2019,Johanssen2019} and requirement elicitation \citep{Liang2015}. No wonder  feedback  is catching on among one-off developers \citep{fitzgerald2017continuous}. Unfortunately,  feedback has not received the same attention for variant-rich systems in general and Software Product Lines (SPLs) in particular \citep{Dakkak2021a}. 

Software Product Line Engineering (SPLE) aims to support the development of a whole family of software products (aka variants) through systematic reuse of shared assets (aka the SPL platform) \citep{Clements2002}. To this end, SPLE distinguishes between two interrelated processes: Domain Engineering (DE), which builds the SPL platform, and Application Engineering (AE), which consists of deriving individual products from the SPL platform. 
In this setting, experiences with  feedback are very rare  \citep{Dakkak2021a}.

Traditionally, feedback is conducted at the application level  \citep{Oordt2021}.  Yet, a hallmark of SPLE is  to move development efforts as much
as possible from AE to DE. Feedback should be no exception. Thus, we advocate for feedback practices to be part of DE and, as such, to be moved upwards from products to the SPL platform. This leads to platform-based feedback rather than product-based feedback.
Unfortunately,  the limited support for continuous deployment  among variability managers (e.g., \textit{pure::variants} \citep{Beuche2019}, \textit{GEARS} \citep{Krueger2018}) hinders this issue from being  grounded on empirical evidence. On these premises, we tackle two research questions:
\begin{itemize}
\item How could `platform-based feedback' be incorporated into SPLE practices?  
\item How would  variability managers be enhanced to support `platform-based feedback'?  
\end{itemize}

In addressing these questions, we contribute by
\begin{itemize}
    \item elaborating on general requirements for leveraging variability managers with feedback services (Section \ref{sec:design-principles});
    \item providing proof-of-concept of how these requirements can be realized  for \textit{pure::variants}  as the variability manager and \textit{Google Analytics (GA)} as the usage tracker (Section \ref{sec:proof-of-concept});
    \item providing proof-of-value through a hybrid evaluation:   a Technology Acceptance Model (TAM) evaluation is first conducted among employees of \textit{pure-systems GmbH} (n=8), followed by a  focus group for the most divergent TAM questions (n=3) (Section \ref{sec:evaluation}).
\end{itemize}

The research is conducted as a  \textit{Design Science Research (DSR) } practice  \citep{Sein2011}. DSR  aims to solve or at least explain the issues of a problematic situation (i.e.,  implicit feedback  not being conducted at the SPL platform level) by favoring a bottom-up approach where researchers join practitioners in their  local problems/solutions, which are next argumentatively generalized. This bottom-up approach fits our research setting where, to the best of our knowledge, no experiences, let alone theories, have been reported on implicit feedback in SPLs  \citep{Dakkak2021a}.  We start with a brief about implicit feedback in one-off development. 

\section{Background: Implicit Feedback}\label{sec:background}

The industry is embracing agile and continuous methods, 
which inherently includes customers' feedback in the analysis loop of the development process \citep{Schon2017,Johanssen2019}. Specifically, the literature distinguishes two customer feedback types: explicit and implicit. The first refers to what users directly say or write about a software (e.g., tweets, reviews written in an app store, etc.) \citep{Oordt2021,Johanssen2019}. By contrast, implicit feedback  automatically collects data about the software usage or execution from which user preferences and trends can be inferred \citep{Maalej2009,Oordt2021}. 
In short, explicit feedback is how users see the software, while implicit feedback depicts how they actually use it.  

Uses of implicit feedback in one-off development include:  
\begin{itemize}
    \item \textit{Requirement Prioritization.} Based on actual feature usage, managers prioritize which requirement implementation or maintenance should be performed first  \citep{Johanssen2019,Hoffmann2020,Oordt2021,JohanssenK2018}.
    \item \textit{Requirement Elicitation.} Discovery of new requirements by analyzing the different usage paths of the users \citep{Hoffmann2020, Oriol2018}
    \item \textit{Optimization.} Improvement of different quality indicators \citep{Martinez-Fernandez2019, Johanssen2019} or optimization of functionalities to be easier and faster for the users to perform them \citep{Oordt2021}.
    \item \textit{Bug Awareness.} Early identification and fixing of bugs by capturing them through crash reports and raised errors in the logs \citep{Oordt2021,Johanssen2019,OlssonB13}.
    \item \textit{User Understanding.} Understanding how users interact with the software (e.g., analyzing how they adapt to new functionalities)\citep{Oordt2021,Johanssen2019}.
\end{itemize}
SPLs certainly need to prioritize requirements, optimize code, or  understand users.  Yet, the fundamental difference rests on the level at which the decision-making process occurs: the platform level rather than  the product level. In particular, in SPLE the prioritization of requirements or the optimization of code  is often conducted at  the platform level rather than based on individual products. Though application engineers  can provide information about specific products, such as usage or bugs, the ultimate decision of which code to optimize or which requirement to prioritize often lies with the Domain Engineering board.  Therefore, we advocate for  implicit feedback to be harmonized with SPLE practices and hence, be moved to the platform level.

\section{Practical case: Implicit Feedback at WACline}\label{sec:wacline}
\begin{figure*}[ht!]
    \centering
    \includegraphics[width=0.9\textwidth]{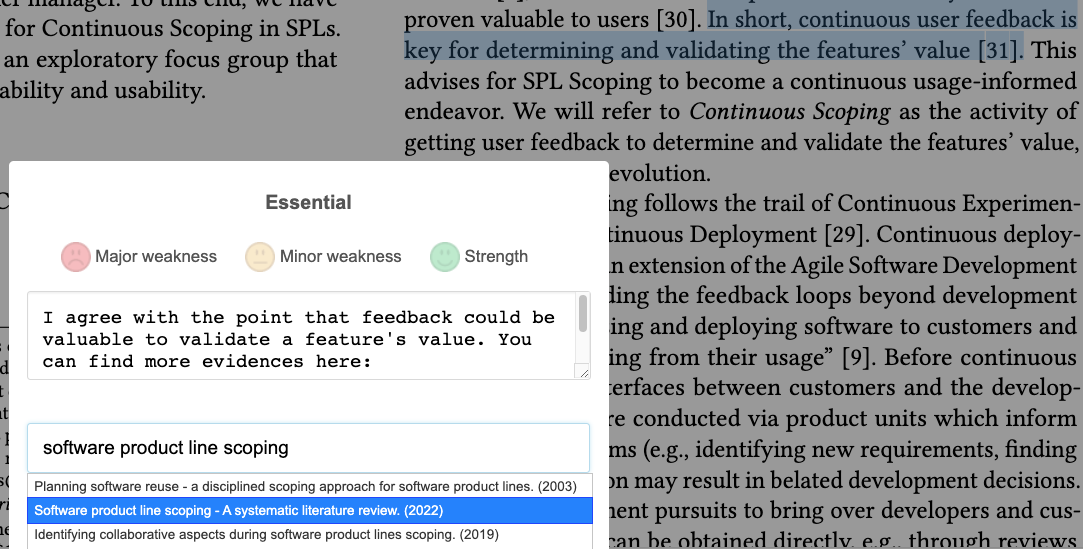}
    \caption{The \textit{Commenting} feature. Besides adding notes, reviewers can include bibliographical references from Google Scholar.}
    \label{fig:CommentingScreenshot}
\end{figure*}

This section introduces the case that triggered this research: \textit{WACline}, an SPL in the Web Annotation domain \citep{Medina2022}. With 85 distinct features, \textit{WACline} supports the creation of custom browser extensions for Web annotation with a current portfolio of eight variants. 
WACline engineers need to track feature usage. The discomfort of supporting  implicit feedback  by directly acting upon the variant's code causes need of a more fitting approach. 

\subsection{The running example}

As the running example, Fig. \ref{fig:CommentingScreenshot} shows the case of the \textit{Commenting} feature.  This feature permits  adding a comment to a highlighted text on the web. In addition, it allows reviewers to query Google Scholar directly without opening a new browser tab. 

\begin{figure*}[ht!]
    \centering
    \includegraphics[width=\textwidth]{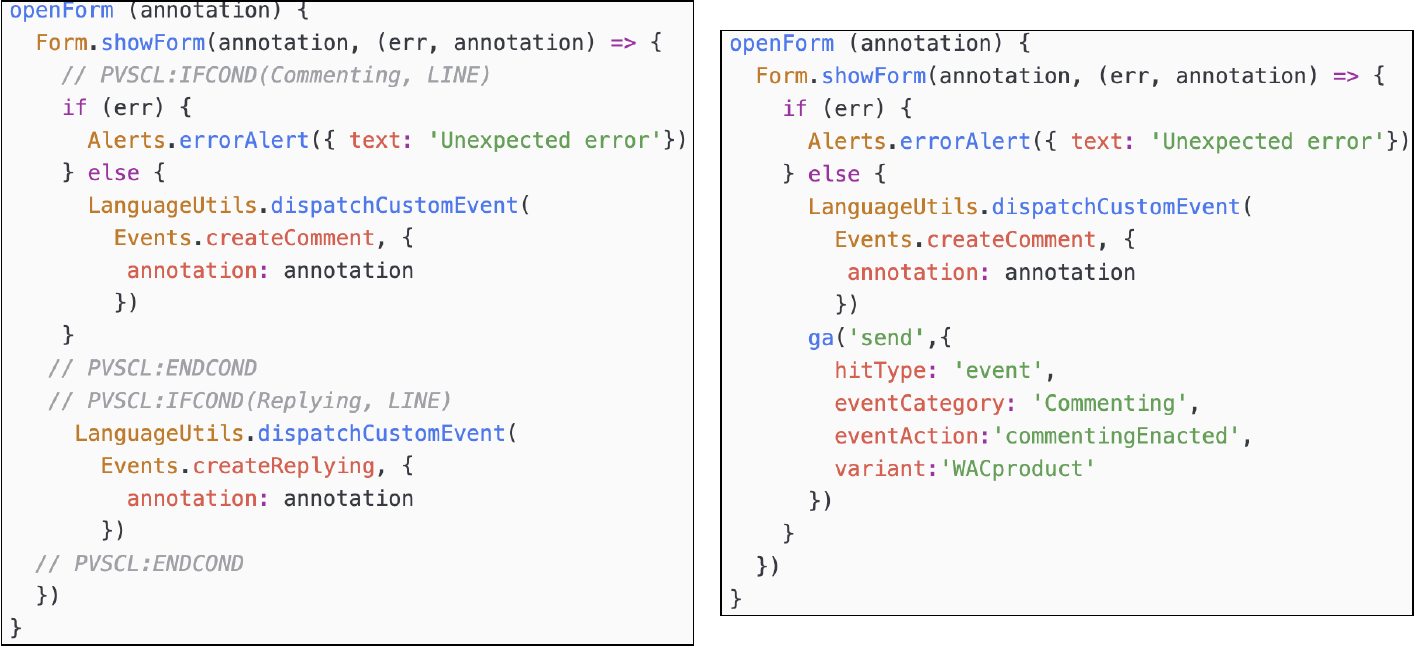}
    \caption{\textbf{Platform} code (with pre-compilation directive) vs Feedback-minded \textbf{variant} code (i.e.,  each execution of this code populates the feedback log with an event occurrence described along with  the  `eventCategory' (e.g., \textit{Commenting}), the `eventAction' (e.g., \textit{CommentingEnactment}) and the `variant' (i.e., the product which raises the event). }
    \label{fig:snippets}
\end{figure*}

Implementation-wise, WACline follows an annotation-based approach to SPLs. This implies that variations are supported through pre-compilation directives that state when a block code is to be included in the final product based on the presence or absence of a feature selection at configuration time. Using  \textit{pure::variants} as the event manager,  WACline variations are denoted through pre-compilation directives that start with an opening directive \textit{//PVSCL:IFCOND} and end with a closing directive \textit{//PVSCL:ENDCOND}.  Fig. \ref{fig:snippets} (left) provides an example with the\textit{ //PVSCL:IFCOND} directives  (aka \textit{\#ifdef} block) framing part of the codebases of  \textit{Commenting}  and \textit{Replying} features.

WACline delivers Web products. Web applications are one of the earlier adopters of implicit feedback. This is achieved with the help of web analytic services such as GA  \citep{w3techs}. GA tracks and reports website usage.  It provides insights into how visitors interact with a website, including the number of visitors, their behavior, and their demographics. Operationally, interaction tracking is achieved  by inlaying tracking instructions (known as `hits' in the GA parlance). Fig. \ref{fig:snippets} (right) displays an example of \textit{Commenting} once incorporated into a \textit{WACproduct} (notice that the \textit{\#ifdef} block is no longer there). \textit{WACproduct} is now `feedback-minded', i.e., each execution of this code populates the feedback log with an event occurrence.  Those hits run in the web browser of each client, collecting data and sending it to the feedback log. Eventually, this feedback log is consulted  through appropriate dashboards \citep{ga-hit}.  

However, \textit{WACproduct} is not just a standalone product, but a variant derived from the WACline platform. Realizing implicit feedback in the conventional way (i.e., at the variant level) causes the `gut feeling' that a fitting approach was needed.

\subsection{The  `gut feeling'}
While providing tracking code at the variant level, WACline engineers were aware of  eventual risks in terms of delays, partially, and chances of `merge hells' in the Git repository. The following paragraphs delve into these issues.

\textit{Risk of delays}. Continuous deployment brings  developers and customers closer, reducing the latency in development decisions often associated with the traditional model where product units serve as the interface between customers and development teams. SPLE introduces an additional layer of indirection between users and code developers. As code developers, domain engineers might not have direct access to the users through the application engineers. This additional layer can lead to further delay in receiving user feedback, slowing down the decision-making process.

\textit{Risk of partiality}. Eager to promptly content their customers, application engineers might be tempted to conduct   feedback analysis locally (i.e., usage tracking limited to their products), favoring changes to be conducted at the product level, overlooking the  implications for the  SPL platform as a whole.

\textit{Risk of `merge hell'}. In SPLE, changes made to the product branches tend to be merged back into the platform branch. If GA hits are added at the application level, then each AE team will inject them for their purposes. However, this way of conducting feedback analysis increases the likelihood of encountering merge conflicts while merging variant branches back into the platform's Master branch.

Alternatively, WACline engineers  attempted to define `feedback' as a feature, but it did not work either. Usage feedback  is not a  "prominent or distinctive user-visible aspect of a variant" but a measurement for decision-making that might involve tracking distinct single features or feature combinations (e.g., Is F1 and F2 used together?). Feedback directives can be considered crosscuts on the SPL's Feature Model.  This situation highlighted two significant inconsistencies when conducting feedback at the variant level:
\begin{itemize}
    \item a conceptual mismatch (tracking subject). SPLs reason in terms of \textit{features}, while GA (and other analytic services) consider products  the unit of tracking,
    \item a process mismatch (scope). SPLs set the analysis  for a collection of products (i.e., the SPL portfolio), while GA is thought to track  a single product.
\end{itemize}

These observations prompted the endeavor to integrate implicit feedback into existing variability management systems, ensuring a platform-wide coverage of feedback.

\section{Requirements for feedback-minded variability managers}\label{sec:design-principles}
\begin{figure*}[ht!]
    \centering
    \includegraphics[width=\textwidth]{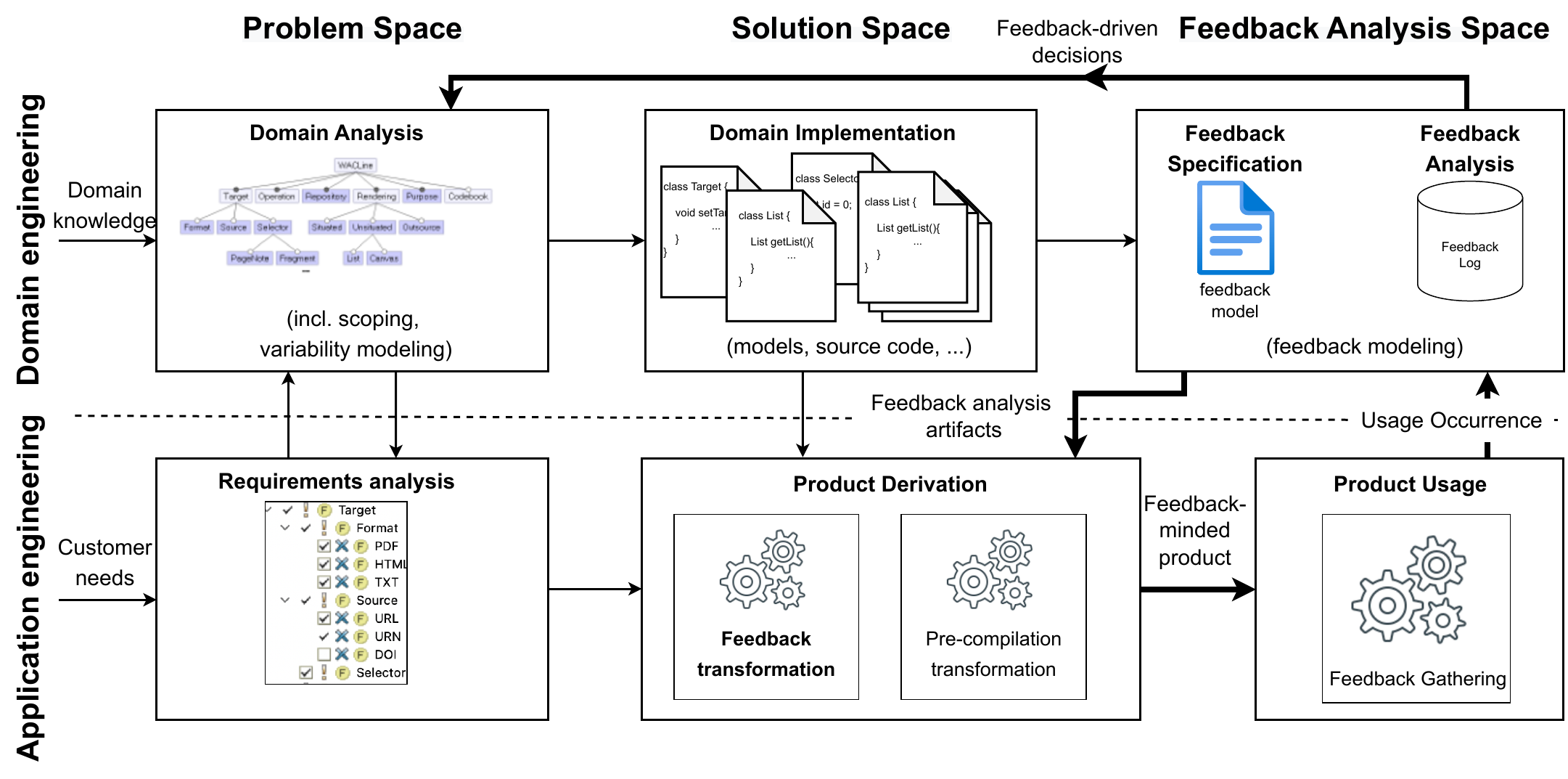}
    \caption{Leveraging SPLE with  Feedback Analysis. The bold line denotes the feedback workflow: analysis-specification-coding-dataGathering-analysis.}
    \label{fig:SPL-overview}
\end{figure*}
Design principles (aka general requirements)  provide guidelines for designing effective solutions that can solve real-world problems and meet the needs of the stakeholders \citep{Gregor2020}. Therefore, design principles are very much framed by the problems and stakeholders where the principles were born.  This section introduces design principles derived from our experience handling implicit feedback in WACline. 

\subsection{General Requirements}

This subsection identifies feedback tasks to be considered during the SPLE process, and hence, they are amenable to being accounted for by variability managers. Fig. \ref{fig:SPL-overview} depicts the traditional SPLE process \citep{Apel2013}  enriched with feedback tasks. Specifically:
\begin{itemize}
\item Domain Engineering now includes  \textit{Feedback Specification} (i.e., the description of the data to be collected in the Feedback log) and \textit{Feedback Analysis} (i.e., the study of the  data in the Feedback log), 
\item Application Engineering now incorporates \textit{Feedback Transformation} (i.e., inlaying feedback tracking instructions during variant derivation) and \textit{Feedback Gathering} (i.e., collecting implicit feedback during variant execution). 
\end{itemize}
 The following subsections delve into these tasks.

\begin{figure*}[th!]
    \centering
    \includegraphics[width=0.85\textwidth]{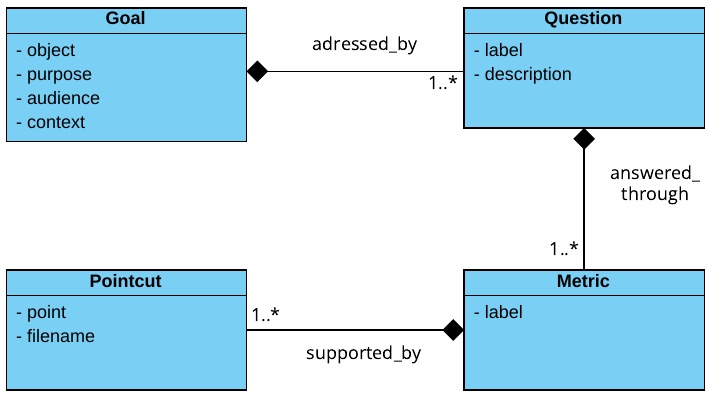}
    \caption{The Feedback Model.}
    \label{fig:metamodel}
\end{figure*}

\subsubsection{Feedback Specification}\label{sec:feedback-specification-req}

Like in one-off development, feature usage can become a main informant for different decision-making issues throughout the lifetime of the SPL platform.  The Goal-Question-Metric (GQM) approach is a popular framework for driving goal-oriented measures throughout a software organization \citep{basili1988tame}. By defining specific goals and identifying key performance indicators, the GQM approach provides a structured way for decision-making. We frame Feedback Specification as a GQM practice with metrics related to feature usage. Fig. \ref{fig:metamodel} depicts our (meta)Model. 

\citet{basili1988tame} introduce the following template for goal description: 
\begin{quote}
    Analyze \textit{<object of study>} with the purpose of \textit{<purpose>}
with respect to \textit{<quality focus>} from the point of view of the \textit{<perspective>} in the context of \textit{ <context>}
\end{quote}

In one-off development, the `object of study' tends to be a single product.  However, SPL heavily rests on features as a significant construct throughout the SPLE process. Accordingly, it is natural  to introduce features as the `object of study' of feedback analysis. After all, chances are that agendas, organizational units or release schedules are set regarding features \citep{Hellebrand2017,Wnuk2009}. Yet, for some analysis scenarios, features might turn too coarse-grained, and  analysis might need to be conducted regarding particular feature combinations (e.g., as described in the pre-compilation directives  that conform the  \textit{\#ifdef} blocks).  Hence, the codebase to be analyzed may be denoted by a single feature or a combination of features. This will identify `the slice of the SPL codebase' that stands for  the object of study.  

Next, this code slice is to be analyzed in the pursuit of distinct purposes (e.g., refactoring, testing, etc.) concerning `usage' from the point of view of a given audience (e.g., domain engineers, testers, product units, product engineers) in a given context. 
Finally, a \textit{Goal} is refined in terms of \textit{Questions}. A question should be open-ended and focus on the specific aspect of the goal that needs to be measured. Hence, questions are to be answered with the help of \textit{Metrics}, i.e., measures that indicate progress towards the goal. 

For example,  consider the \textit{Commenting} feature as the `object of study' (see Fig. \ref{fig:CommentingScreenshot}). For this object, possible purposes  include: 
\begin{itemize}
\item feature scoping. Engineers might want to assess the extent Google Scholar is queried from within \textit{Commenting}. The rationale goes as if this particular utility is seldom used w.r.t. the usage of \textit{Commenting}, then it might be better to detach it from \textit{Commenting} and offer it as a separate sub-feature.
\item feature optimization. \textit{Commenting} is made available in two ways: `double click' and `right click'. Engineers wonder whether this double choice may confuse users and, if so, which is the most popular approach for this feature to show up.
\end{itemize}
If the SPL Control Board (i.e., the committee  in charge of the SPL fate) conducts this study in the context of planning the next WACline release, the resulting goal is left as follows:
\begin{quote}
Analyze \textit{Commenting} with the purpose of \textit{scoping \& optimization}
with respect to \textit{usage} from the point of view of the \textit{Control Board} in the context of \textit{the planning for the next release}
\end{quote}
The example so far assumes the context to be the SPL platform as a whole. Though the object of the study is limited to the Commenting's codebase (i.e., the \textit{\#ifdef} blocks involving Commenting), the context  is the entire SPL portfolio, i.e., track Commenting no matter the variant in which this feature is deployed. 

In addition, we also came across scenarios where the analyst was interested in the usage of a feature but for a particular set of variants. Back to the running example,   \textit{Commenting} adds  menus and widgets to the Web interface.  If utilized with other GUI-intensive features, such as \textit{Highlighting}, \textit{Commenting} may lead to cluttered interfaces that can potentially discourage users. It should be noted that \textit{Commenting} and \textit{Highlighting} have no feature dependency; hence, they do not appear together in pre-compilation directives. However, the combining effect regarding the cluttered interface  requires the tracking of  \textit{Commenting} for those variants where Highlighting is also deployed.
We underline that `contextual features' might exhibit no dependency on  `objectual features' as captured in feature models yet exhibit some sort of implications that qualify the context of the tracking (e.g. F1's power consumption might impact the response rate of F2 and F3, yet this implication might not always be reflected in the feature model). In this case, the goal would look as follows:
\begin{quote}
Analyze \textit{Commenting} with the purpose of \textit{usability evaluation}
with respect to \textit{usage} from the point of view of the \textit{UX designer} in the context of \textit{Highlighting}
\end{quote}
Here, the  \textit{Highlighting} feature characterizes the context where Commenting is to be tracked, i.e., the subset of the SPL portfolio subject to tracking.

We now move to the questions. Previous goals on Commenting  could be addressed through two questions with their respective metric counterparts,  namely: 
\begin{itemize}
\item How is \textit{Commenting} used? Commenting permits querying Google Scholar from within. The Control Board wonders about the extent   this feature is coupled with the fact of commenting, and if  sparse, decides to offer it as a separate sub-feature. To this end, \textit{GoogleScholarHappened} metric is introduced;
\item How is \textit{Commenting} enacted? This feature offers two enactment modes:  \textit{DoubleClickEnactment} or \textit{RightClickEnactment}. Assessing the popularity of each of these modes helps to determine if they are alternatives or if any of them is residual, and hence, a candidate to step out in the next release.
\end{itemize}

For these metrics to be enactable, the feedback specification needs to indicate how is going to be supported: the pointcut. A pointcut is a join point which denotes a given line of code within a file. A pointcut denotes the point in  `the slice of the SPL codebase'  (i.e., the object)  where a  tracking hit should be raised and collected in the feedback log. A metric might involve one or more pointcuts. As we will see later, the metric \textit{DoubleClickEnactment} will require two pointcuts, one for each of the points in the code where double-click can be enacted.  Section \ref{solution:feedback-spec} will introduce the concrete syntax for this model using YAML. 

\subsubsection{Feedback Transformation}
\begin{figure*}
    \centering
    \includegraphics[width=\textwidth]{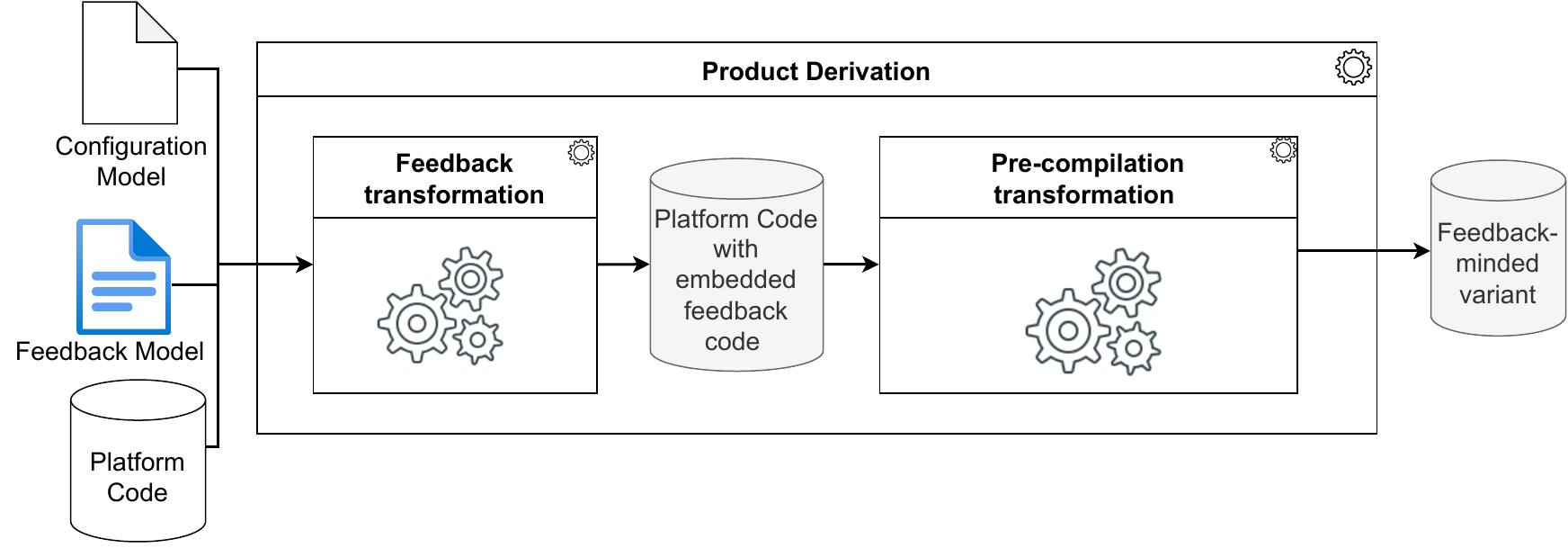}
\caption{Two-step Product Derivation. First, the Feedback Transformation injects the tracking hit along with  the feedback model. Second, the Pre-compilation Transformation filters out those \textit{ifdef} blocks that do not meet the configuration model. The result is a feedback-minded product. }
    \label{fig:product-derivation-theory}
\end{figure*}

Feedback transformation refers to the interpretation of the feedback model. 
In one-off development, the analyst manually introduces the tracking instructions into the product codebase. SPLs aim at reducing, if not removing, coding during AE. The  SPL way is to resort to `transformations' that generate the variant through `transforming' the SPL platform along with a given configuration model.
Likewise, feedback goals might be better supported through `transforming' the SPL platform rather than directly acting upon the variant code. Rather than polluting the platform code, `separation of concerns' advises for   feedback-specific code to be added on the fly during Product Derivation.  Specifically, Feedback Transformation takes the SPL platform, the configuration model and the feedback model as input and injects the corresponding tracking code (see Fig. \ref{fig:product-derivation-theory}). This code is next subject to the traditional transformation based on pre-compilation directives. The resulting variant will be feedback-minded, i.e., it will populate the feedback log at run-time.

\subsubsection{Feedback Gathering}
Feedback gathering refers to capturing usage data from users \citep{Johanssen2019}. Traditionally, in the same vein as any other data, usage data should meet specific quality properties \citep{Redman13}, namely accuracy (i.e., tracking hits should be error-free and reflect real-world values), completeness (i.e., tracking hits should include at least all the information specified in the feedback model), consistency (i.e., data gathered should be consistent among the different products) and, timeliness (i.e., data should always be up-to-date and reflect the real use of the platform). Thus, feedback gathering goes beyond a programming issue to also include data governance considerations.  The question is what  SPLE brings to account for these quality factors: 
\begin{itemize}
    \item accuracy. SPLE-wise, the feedback specification and specifically, the contextual features provide a declarative way to set the scope of the feedback analysis. Rather than naming the products subject to the tracking (extensional approach), SPLs offer features as an implicit means to single out the variants of interests.  On the downside, faults in the feedback specification will have a more significant impact as they propagate to a larger set of variants;
    \item completeness. SPLE-wise, completeness imposes a careful consideration not only of the contextual features to be considered but also of the end-users (or  framing installations) where the variants are to be deployed. This might be an issue if domain engineers do not have access to the ultimate users/installations where the variants are run; 
    \item consistency. SPLE-wise, the transformation approach ensures that the feedback code is generated in the same way no matter  the variants in which this code ends up being injected; 
    \item timeliness. SPLs evolve quicker than one-off applications if only by the number of variants they consider and the different sizes of the codebase. The larger the number of variants and the codebase, the larger the chances for feedback analysis to evolve, and hence, the more challenging to keep usage data up-to-date. Traditional product-based feedback does not scale to this volatile scenario. Keeping pace with feedback analysis evolution  calls for  model-driven approaches where analysis is described at a higher-abstraction level, and transformation takes care of implementation details. Model-driven not only accounts for code consistency but also speeds up development, hence, reducing the cost of updating feedback specification which facilitates up-to-date data.
\end{itemize}

\subsubsection{Feedback Analysis} 
To aid decision-making  in dynamic environments, dashboards are deemed essential \citep{Lopez2021}. A dashboard is a graphical user interface that provides an overview of key performance indicators, such as feature usage, relevant to a particular business objective or process (such as feature optimization) \citep{Few2006}. \textit{Google Analytics} offers dashboards for website owners to analyze trends and patterns in their visitors' engagement with the website \citep{Plaza2011}. However, SPL scenarios differ  in two key aspects. Firstly, the focus of the tracking is not a single application but rather (a subset of) the SPL portfolio. Secondly, features become the central unit of analysis. Thus, dashboards must be designed to accommodate these unique characteristics.

\subsection{Validating the requirements }\label{sec:IF4SPL-eval}
\begin{figure*}
    \centering
    \includegraphics[width=\textwidth]{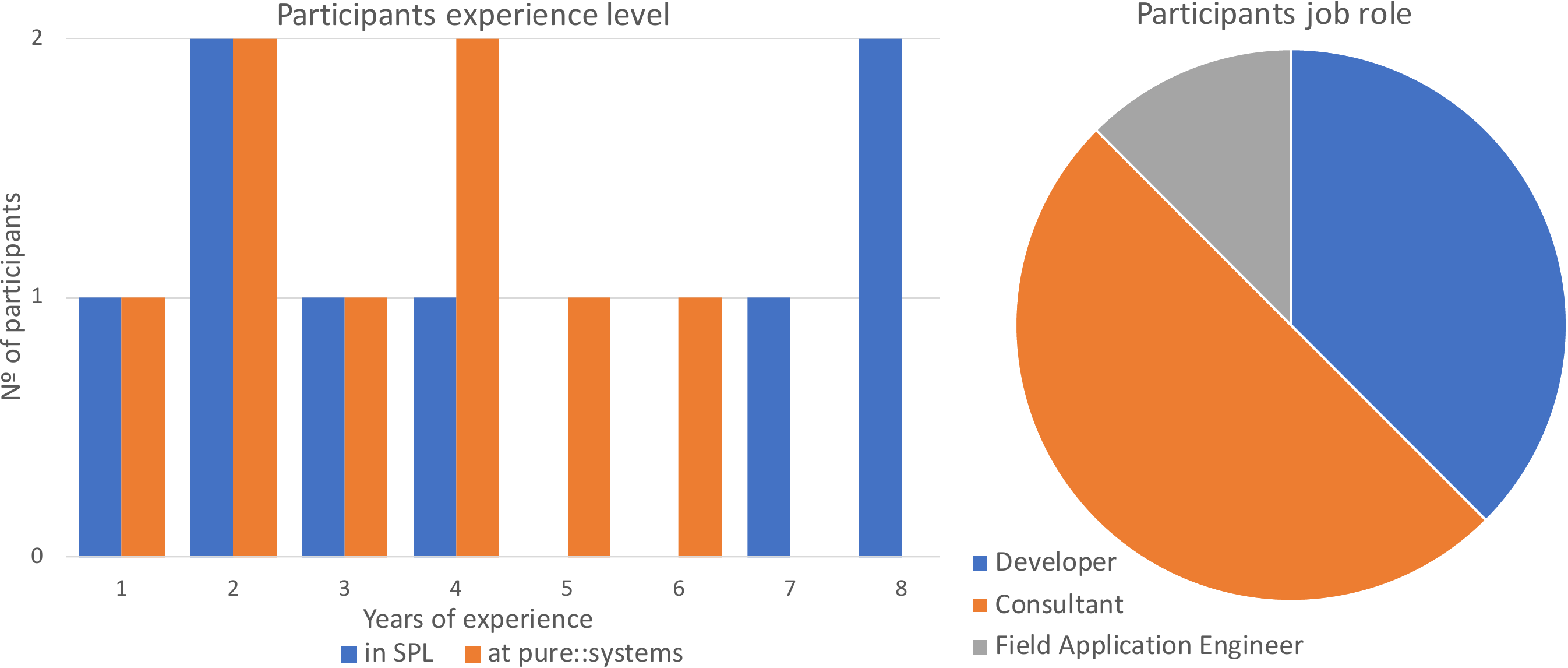}
    \caption{Participants demographics: SPL experience \& role played. }
    \label{fig:demographics}
\end{figure*}

\begin{figure*}[th!]
    \centering
    \includegraphics[width=1\textwidth]{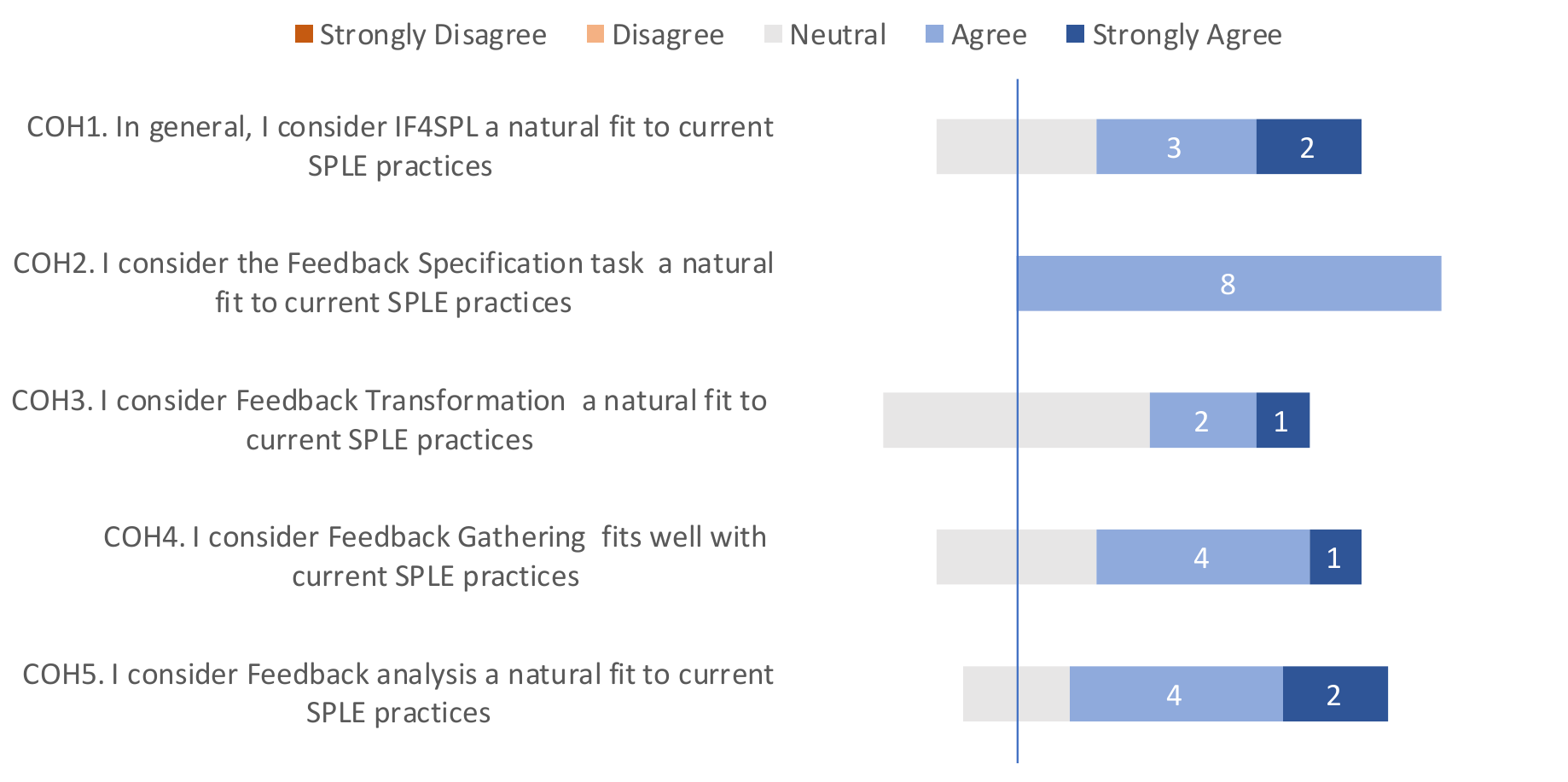}
    \caption{Assessing coherence for IF4SPL. }
    \label{fig:coherence}
\end{figure*}

\begin{figure*}[th!]
    \centering
    \includegraphics[width=1\textwidth]{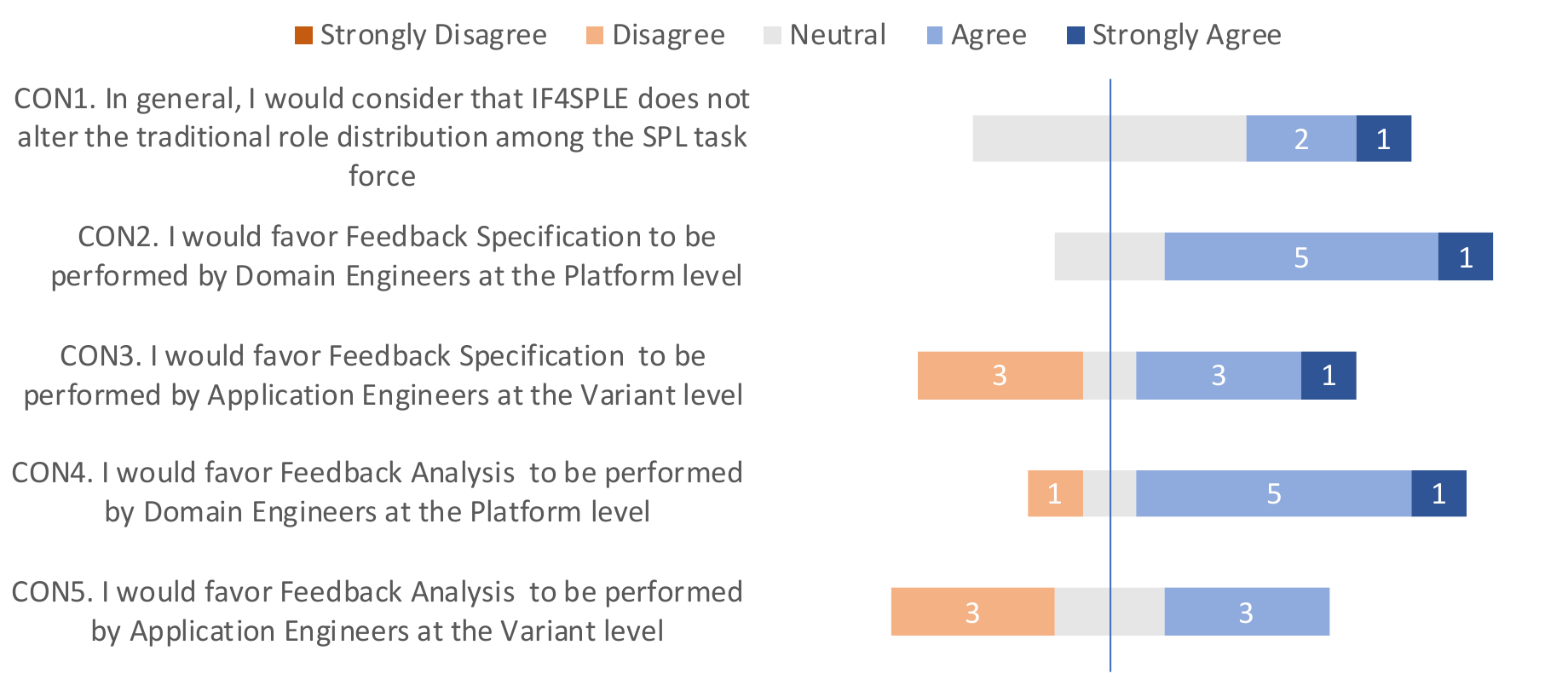}
    \caption{Assessing consistency  for IF4SPL.}
    \label{fig:consistency}
\end{figure*}

The previous subsection introduced a way to support implicit feedback in SPLE, named IF4SPLE (i.e., Implicit Feedback for SPLE).
This proposal was derived from our experience handling implicit feedback in WACline. To improve external validity, this section conducts a survey on the extent IF4SPLE accounts for  \textit{coherence} (i.e., the quality of forming a unified whole with traditional SPLE practices) and \textit{consistency} (i.e., the quality of seamlessly fitting with prior roles in SPLE). The next paragraphs describe the evaluation of IF4SPLE principles. 

\paragraph{Goal} The purpose of this study is to \textit{assess} the goodness of the  IF4SPL principles for \textit{seamlessly introducing Implicit Feedback into SPLE practices} from the point of view of \textit{annotated-based SPL practitioners} in the context of  \textit{pure-systems GmbH}\textit{pure-systems GmbH}\footnote{\url{https://www.pure-systems.com/}}   and annotated-based SPLs.

\paragraph{Participants} To look into the generality of the solution, we resorted to  \textit{pure-systems GmbH}, which is a leading provider of software product lines and variant management engineering tools, including \textit{pure::variants}. The suitability of this company is  twofold. First, \textit{pure-systems GmbH} is well placed to assess the extent \textit{FEACKER} is seamlessly integrated with its own tool (i.e., \textit{pure::variants}). Second,  \textit{pure-systems GmbH} acts as a consultant for a broad range of companies, and hence, faces different requirements and domains. Hence, its participation will provide a high variation of settings in which to support generalization.  

\paragraph{Design} The experiment was conducted within the monthly seminar regularly kept at \textit{pure-systems GmbH}. The seminar lasted for 90 minutes. The participants were introduced to the importance of Implicit Feedback in single-off development and the interest in bringing these benefits to SPLE. Next, IF4SPLE was introduced. At this time, participants were asked to fill up the questionnaire (see next paragraph). Fifteen employees attended the seminar, yet only eight completed the survey. For these eight employees, demographic data  is displayed in Fig. \ref{fig:demographics}. 


\paragraph{Instrument}
To better profile what is meant by `a seamless introduction', we characterize this adverb in terms of \textit{coherence} (i.e., the quality of forming a unified whole with traditional SPLE practices) and \textit{consistency} (i.e., the quality of seamlessly fitting with prior roles in SPLE). We resort to an \textit{ad-hoc} questionnaire where items are assessed through a five-point Likert that is refined for each  of the tasks introduced by IF4SPLE (i.e., Feedback Specification, Feedback Transformation, and so on). Whereas coherence looks at the process, consistency regards the roles played by participants in an SPL organization. Specifically, we consider two roles (i.e., DE vs AE). The rationale reads as Implicit Feedback might be conducted as part of DE with the participation (or not) of application developers. And the other way around, it is possible for Implicit Feedback to be defined for a set of products without implying the whole platform. Figs. \ref{fig:coherence} and \ref{fig:consistency} show the results of IF4SPLE's coherence and consistency, respectively.

When it comes to assessing the internal consistency, or reliability, of questionnaires,  Cronbach’s alpha coefficient measure is used. It helps determine whether a collection of questions consistently measures the same characteristic. Cronbach’s alpha quantifies the level of agreement on a standardized 0 to 1 scale. Higher values indicate higher agreement between items. Thus, high Cronbach’s alpha values indicate that response values for each participant across a set of questions are consistent. On these grounds, we  calculate the Cronbach’s alpha values of the questionnaire items for  coherence and consistency, leading to a result of   0.63, and 0.8 alpha values, respectively. In the case of consistency, we reversed the results of questions CON3 and CON5 as they favor AE to conduct implicit feedback affairs. An acceptable degree of $\alpha$ reliability is defined as 0.6 to 0.7, and a very good level is defined as 0.8 or greater \citep{Ursachi2015}. Therefore, we can consider the questionnaire reliable enough.

\paragraph{Results}

\textbf{Coherence.} Participants generally believe IF4SPLE is a natural fit for current SPLE practices (question COH1). Feedback Specification (question COH2) and Feedback Analysis (question COH5) were the tasks that unanimous agreement reached about being conducted at the level of DE, hence moving Implicit Feedback from being a product concern to being a platform concern. Surprisingly, while positive, feedback transformation (i.e., injecting feedback concerns when the product is generated mimicking traditional techniques based on pre-compilation directives) did not meet our expectations, with most participants ranking it as `neutral'.  

\textbf{Consistency.} Participants generally considered that IF4SPLE does not alter the traditional role distribution among the SPL task force (question CON1). That said, in line with IF4SPLE,  \textit{pure::variant}'s employees seem to favor domain engineers before application engineers when conducting feedback specification and analysis (questions CON2 and CON4).


The next section looks at the feasibility of this approach, using  \textit{pure::variants} as the variability manager.

\section{Proof-of-concept: implicit feedback in \textit{pure::variants} \label{sec:proof-of-concept}}
\begin{figure*}[th!]
    \centering
    \includegraphics[width=\textwidth]{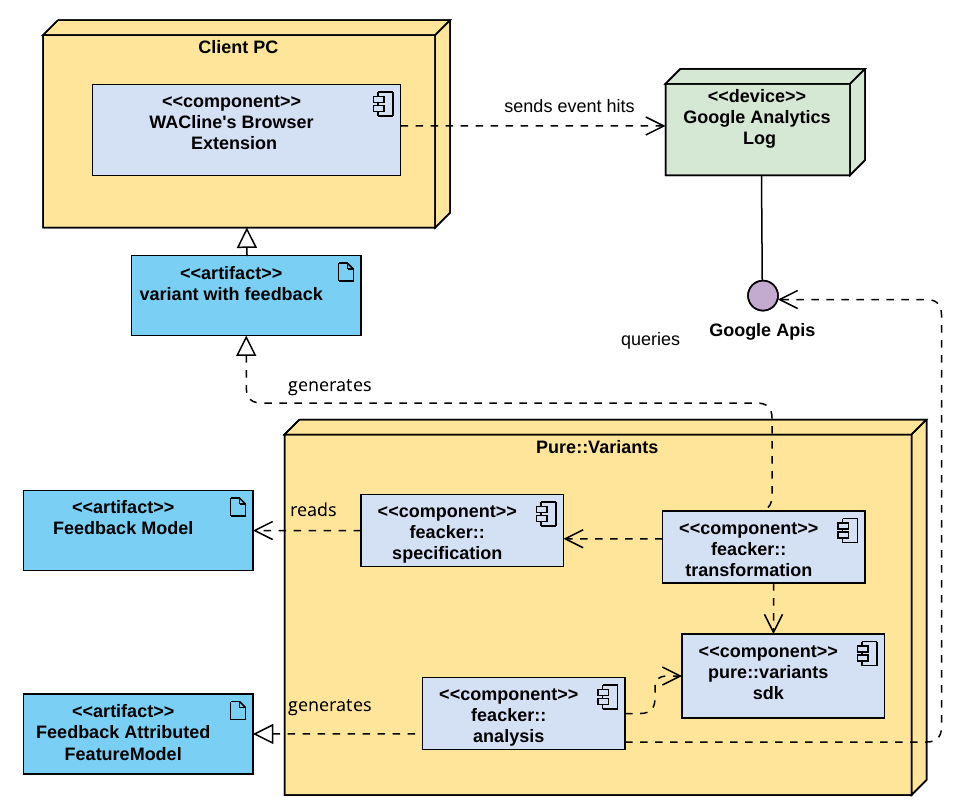}
    \caption{\textit{FEACKER}'s deployment diagram. \textit{Pure::variants} is extended with three new components: \textit{feacker::specification, feacker::analysis} and\textit{ feacker::transformation}}.
    \label{fig:FEACKER-deployment}
\end{figure*}

This section introduces \textit{FEACKER} (a portmanteau for FEAture and traCKER), an extension for  \textit{ pure::variants} along the IF4SPLE principles. \textit{FEACKER} is publicly available at \url{https://github.com/onekin/FEACKER}.

 The deployment diagram of the \textit{FEACKER} system is presented in Fig. \ref{fig:FEACKER-deployment}. The specification and transformation components are responsible for incorporating the tracking code, utilizing GA as the tracking framework. At run-time, variant execution on the \textit{Client PC} generates data in the GA log, which is later analyzed by the analysis component of \textit{FEACKER}. This component processes the feedback data and presents it through an attributed feature model, as described in Section \ref{sec:fea-analysis}\footnote{The implementation of the solution relies on the \textit{pure::variants} SDK, utilizing three extension points: specification, transformation, and analysis. It should be noted that \textit{pure::variants} uses pre-compilation directives for variability support and introduces the concept of a Family Model, which separates the SPL platform implementation through logical names referred to as `model elements'.}. Details follow.

\subsection{Feedback Specification}\label{solution:feedback-spec}
\begin{figure}[th!]
    \centering
    \includegraphics[width=0.8\textwidth]{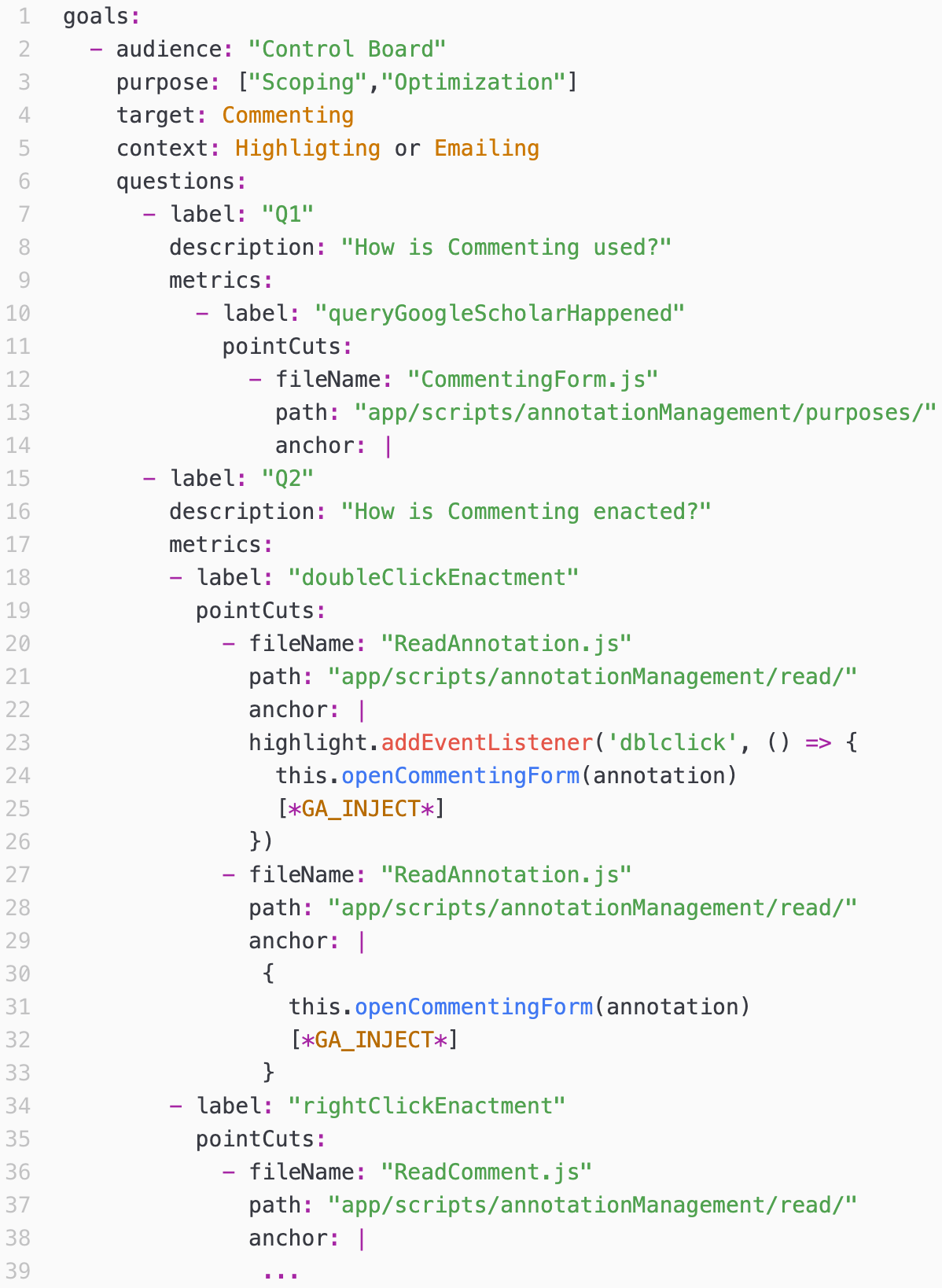}
    \caption{The feedback model for the goal: Analyze \textit{Commenting} with the purpose of \textit{scoping \& optimization}
with respect to \textit{usage} from the point of view of the \textit{Control Board} in the context of \textit{variants that exhibit either Highlighting or Emailing}.}
    \label{fig:feedback-model}
\end{figure}

\textit{FEACKER} resorts to the YAML language for the concrete syntax of the Feedback Model introduced in Section \ref{sec:feedback-specification-req}.  Fig. \ref{fig:feedback-model} provides a  \textit{myFirstfeedback.yaml} snippet for the running example. Clauses on lines 2-5 capture the Goal.  Two clauses have a documentation purpose ( \textit{audience} \& \textit{purpose}), whereas  \textit{target} and \textit{context} have execution implications. Both hold a feature-based predicate  similar to the one in pre-compilation directives (so far, only AND/OR are supported).  \textit{Context} holds `a contextual predicate' to delimit the variants whose configuration model should match this predicate. Target holds `an objectual predicate' to characterize the  \textit{\#ifdef} blocks whose pre-compilation directives should satisfy this predicate. Thus,  the slice of the platform codebase to be tracked is intensively defined by the \textit{\#ifdef} blocks whose pre-compilation directives satisfy the objectual predicate when deployed in variants matching the contextual predicate
subject to the tracking (i.e.,  the slice of the platform codebase to be tracked). Both predicates are consulted during feedback transformation to filter out the \textit{\#ifdef} blocks subject to tracking as a result of matching the variability of interest (object of study) in the appropriate variants (the context). Back to the example, the tracking scope is that of \textit{\#ifdef} blocks whose pre-compilation directives hold \textit{Commenting} when in variants that exhibit either \textit{Highlight} or  \textit{Emailing}. 

As for questions, they do not have any execution implications apart from structuring the metrics. The sample snippet introduced two questions: Q1 and Q2. The former is grounded on \textit{QueryGoogleScholarHappened} metric. As for Q2, it is based on the counting of \textit{DoubleClickEnactment} and \textit{RightClickEnactment}. 

Finally, pointcuts  are captured through three clauses. \textit{FileName} and \textit{Path} single out the code file in \textit{pure::variants}'  Family Model. As for \textit{the anchor}, it holds  the line of code where the GA\_INJECT directive is to be injected at transformation time.

\subsection{Feedback Transformation}
\begin{figure*}[th!]
    \centering
    \includegraphics[width=1\textwidth]{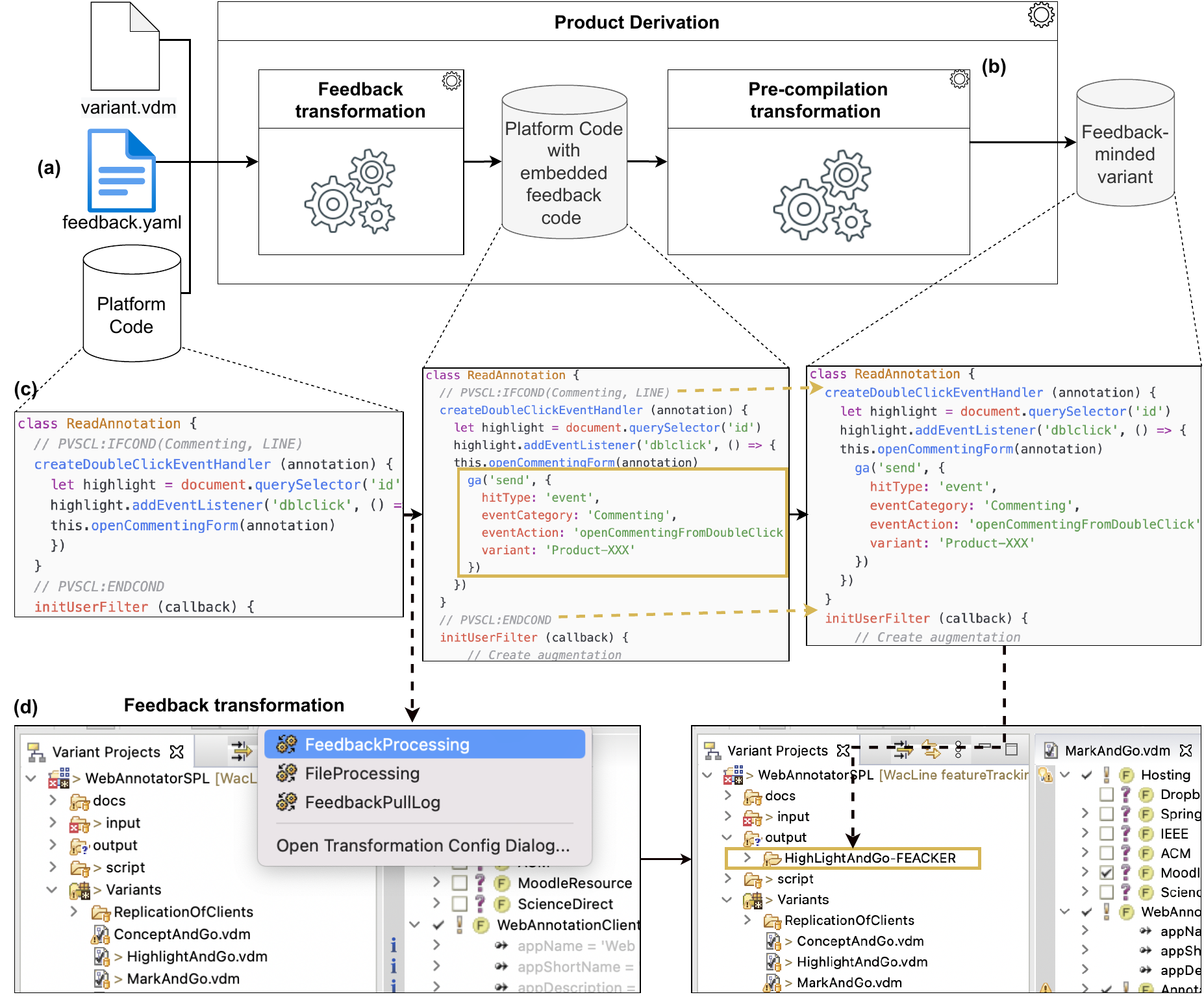}
\caption{FEACKER's realization of Fig. \ref{fig:product-derivation-theory} with the feedback model supported as a \textit{yaml} file (a). The Feedback Transformation takes the platform code as input and injects the GA hits to the corresponding \textit{\#ifdefs} (c). The resulting code is next subjected to the Pre-compilation transformation that filters out   \textit{\#ifdefs}  that do not meet the configuration model.  
GUI-wise,  \textit{pure::variants} is  extended with a second  user interaction: \textit{FeedbackProcessing} (d).  On clicking, FEACKER conducts the Feedback  Transformation, and next invokes \textit{pure::variants}' \textit{FileProcessing} to launch the Pre-compilation Transformation. The result is a feedback-minded variant, i.e., its usage  will populate the feedback log. }
    \label{fig:product-derivation-FEACKER}
\end{figure*}

In \textit{pure::variants}, Product Derivation has two inputs: the configuration model (a .vdm file) and the SPL platform. The output of this process is a variant product, where \textit{\#ifdef} blocks that do not align with the configuration model are removed. This filtering of non-conforming blocks is achieved in \textit{pure::variants} by selecting the `File Processing' option. This is depicted on the right side of Fig. \ref{fig:product-derivation-FEACKER} (b).

In  \textit{FEACKER}, Product Derivation extends \textit{pure::variants'} through an additional step: Feedback Transformation. The output is also a variant, but now the variant code includes GA hints in accordance with the feedback model. 
Fig. \ref{fig:product-derivation-FEACKER} (c) illustrates the progression of a code snippet. The process departs from the platform codebase. The Feedback Transformation injects the GA hits along with the feedback model. GA hits are composed of an interaction (\textit{eventAction}) that happens during the usage of the feature (\textit{eventCategory}) while using a given product (\textit{variant}). Finally, the pre-compilation directives are removed from the code. The output is a feedback-minded variant. Refer to Appendix \ref{ap:algorithms} for the algorithm. 




\subsection{Feedback Gathering}
\textit{FEACKER} resorts to GA for feedback gathering. Event data is collected into the GA feedback log. The utilization of GA presents both advantages and disadvantages. The benefits include GA's ability to comprehensively collect event data regarding visitors, traffic sources, content, and conversions. Additionally, GA provides advanced data visualization tools and it is highly scalable. On the downside, keeping event data at Google might  jeopardize confidentiality.    

\subsection{Feedback Analysis}\label{sec:fea-analysis}
\begin{figure*}[th!]
    \centering
    \includegraphics[width=0.95\textwidth]{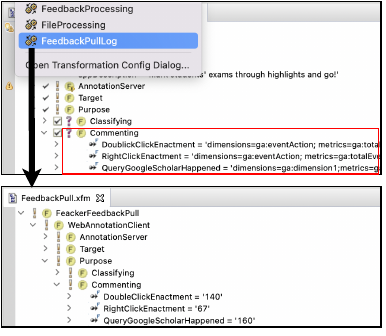}
    \caption{ Attributed Feature models: \textit{Commenting} holds four attributes for GA API calling parameters  (top side). Select \textit{FeedbackPull} from the drop-down menu for these API calls to be enacted. Results are stored in  \textit{FeedbackPull.xfm}, a copy of the input feature model where feature-usage attributes will show the metric values as returned by the API call (bottom side).}
    \label{fig:feedbackpull}
\end{figure*}

GA provides  data visualization tools  to track single apps. By contrast, we pursue feature-centric dashboards that track an  SPL portfolio.  This is a topic in its own right. For completeness' sake, this subsection explores the use of attributed feature models for feedback visualization.

In an attributed feature model, extra-functional information is captured as attributes of features  \citep{benavides2010automated}. Examples include non-functional properties (e.g., cost, memory consumption, price) or  rationales for feature inclusion. This  permits leveraging  the automated analysis of feature models by incorporating these usage parameters \citep{galindo2019automated}. Likewise, we add `usage' as a feature attribute.
However, unlike other attributes such as cost or memory consumption, `usage' is highly volatile and needs frequent updates. In this setting, databases resort to derived attributes, i.e., attributes that do not exist in the physical database but are derived at query time from the database. Regarding derived attributes, two steps are distinguished: definition (when the query is defined)  and enactment (when the query is triggered). 

\textit{FEACKER} supports this vision, namely:

\begin{itemize}
    \item definition time. Analysts might enlarge the feature model with usage attributes.  These attributes hold a GA query to obtain the aggregate for this metric using the GA's API. Fig. \ref{fig:feedbackpull} (top) illustrates this vision for our running example. \textit{Commenting} is the object of study for three metrics: \textit{DoubleClickEnactment, RightClickEnactment}, and \textit{QueryGoogleScholarHappened}, 
    \item enactment time. It is up to the analysts to decide when to enact these queries. To this end,  \textit{FEACKER} adds a `FeedbackLogPull' option to the \textit{pure::variants} processing menu. When selected, \textit{FEACKER} (1) retrieves the queries held as values of the attributes,  (2) executes the queries, and (3) creates a clone of the feature model, where attributes  now hold the GA responses (Fig. \ref{fig:feedbackpull} (bottom)). Refer to Appendix \ref{ap:algorithms} for the algorithm. 
\end{itemize}


At present, query results are integrated into the model without further processing.  Subsequent executions of `FeedbackPull' will replace the existing clone with a new one that mirrors the current feedback log.  All in all, using feature models falls short. First, it is not evident how to visualize metrics that involve multiple features (e.g., usage of F1 but not F2). Second, feature models offer a static view of the current variability,  whereas feedback analysis necessitates a representation of how usage evolves. This calls for \textit{pure::variants} to be enlarged with full-fledged dashboards. Alternatively, extending GA dashboards with feature-centric visualizations. This is still a work in progress.

\section{Proof-of-value: Evaluation}\label{sec:evaluation}

\subsection{Design of the Evaluation}
\begin{table}[ht!]
\centering
\footnotesize
\caption{In the pursuit of a realistic evaluation.}
\begin{tabular}{|l|llc|l|}
\hline
\textbf{}              & \multicolumn{3}{c|}{\textbf{Realistic ...}}                                                            & \textbf{}           \\ \hline
\textbf{}              & \multicolumn{1}{l|}{\textbf{participants}} & \multicolumn{1}{l|}{\textbf{tasks}} & \textbf{environment} & \textbf{Instrument} \\ \hline
\textbf{Pure::Systems} & \multicolumn{1}{c|}{$\checkmark$}                   & \multicolumn{1}{c|}{$\times$}            & $\checkmark$                 & TAM                 \\ \hline
\textbf{WACline}       & \multicolumn{1}{c|}{$\checkmark$}                   & \multicolumn{1}{c|}{$\checkmark$}            & $\times$                 & Focus Group         \\ \hline
\end{tabular}
\label{tab:evaluation}
\end{table}
DSR  emphasizes that authenticity is a more important ingredient  than controlled settings \citep{Sein2011}. For an authentic evaluation, \citet{Sjoberg02} introduce three factors: (1) realistic participants, (2) realistic tasks, and (3) a realistic environment. Sj{\o}berg et al. recognize the difficulties of meeting all factors simultaneously, given the developing nature of the interventions in research.

\textit{FEACKER} is not an exception; in fact, quite the opposite. However, the participation of practitioners from WACline and \textit{pure-systems GmbH}  provides a unique opportunity to assess the utility of \textit{FEACKER}'s outcomes. The practitioners from WACline meet the first criteria of realistic participants, as WACline is web-based. They also used \textit{FEACKER} in a real setting. However, WACline's limited size may limit the results' generalizability.
On the other hand, practitioners from \textit{pure-systems GmbH} are well placed to assess the seamless integration of \textit{FEACKER} with  \textit{pure::variants}. Furthermore, the company is a consultant for a broad range of customers, providing insights on the extent the approach could be generalized to settings other than Web-based variants. However, they have not used \textit{FEACKER} in a real task. 
As a result,  we opt for a hybrid evaluation that combines both  quantitative and qualitative evaluation. Specifically (see Table \ref{tab:evaluation}),

\begin{itemize}
\item for quantitative evaluation, we  resort to the Technology Acceptance Model (TAM) as the instrument and \textit{pure-systems GmbH} as the participant. TAM is founded upon the hypothesis that technology acceptance and use can be explained by a user’s internal beliefs, attitudes, and intentions  \citep{turner2010does},
\item for qualitative evaluation, we draw on Focus Groups as the instrument and WACline practitioners as the participants. Focus groups are suggested in Software Engineering as an empirical method for getting practitioner input and qualitative insights, particularly in the early phases of research for identifying issues \citep{Kontio2008}. In this case, the evaluation resorts to a realistic task (i.e., the Web-based \textit{Commenting} feature), and the 
focus groups are structured along  the dissents expressed in the TAM evaluation.  
\end{itemize}

This hybrid approach rests upon the assumption that both populations (i.e., \textit{pure-systems GmbH}   and WACline) share a common mindset that sustains that  \textit{the intention of use} of the former serves to guide \textit{the real use} of the latter.  This is undoubtedly a threat to `hybrid-ness' validity, yet this likeness between both populations is (partially) supported by both using \textit{pure::variants} as the variability manager.

\begin{figure*}[ht!]
    \centering
    \includegraphics[width=\textwidth]{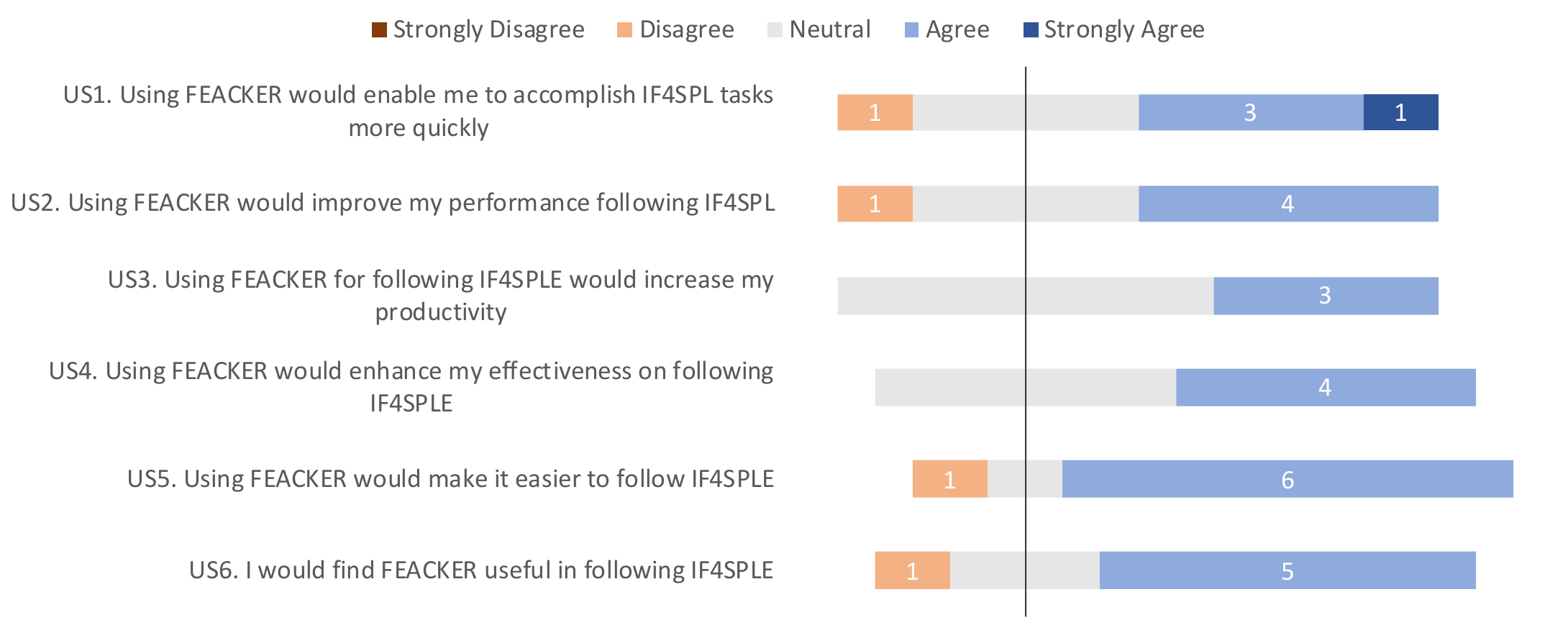}
    \caption{\textit{FEACKER}'s  perceived usefulness. }
    \label{fig:usefulness}
\end{figure*}

\begin{figure*}[ht!]
    \centering
    \includegraphics[width=1\textwidth]{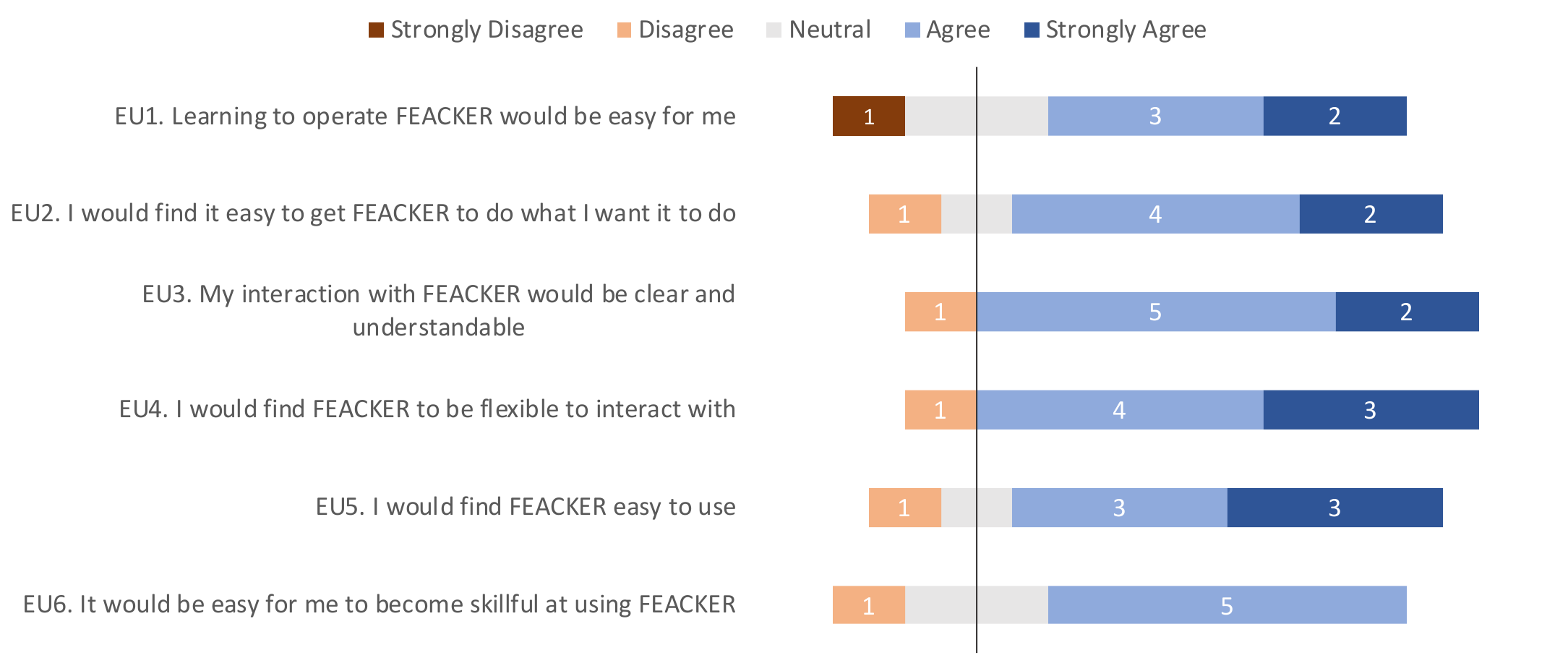}
    \caption{ \textit{FEACKER}'s   perceived ease of use. }
    \label{fig:ease-of-use}
\end{figure*}

\subsection{TAM evaluation}

\paragraph{Goal} The purpose of this study is to \textit{assess} the perceived ease of use and usefulness of \textit{FEACKER} with respect to  \textit{introducing IF4SPLE practices into pure::variants} from the point of view of \textit{annotated-based SPL practitioners}  in the context of \textit{pure-systems GmbH}.

\paragraph{Participants} We sought the assistance of employees at \textit{pure-systems GmbH}. As they were  involved in commercialising and developing \textit{pure::variants}, we believed they were suitably positioned to assess its `ease of use'. Furthermore, half of the participants were consultants, exposing them to a wide range of domains, thereby providing a wider perspective on the FEACKER's `usefulness'.

\paragraph{Design} FEACKER was evaluated as part of the monthly seminar in which the IF4SPLE was previously evaluated (as described in Section \ref{sec:IF4SPL-eval}). The seminar was structured as follows. An initial 90-minute session provided an overview of implicit feedback, including an in-depth examination of the  \textit{FEACKER's} components. A live demonstration of the tool followed this. Participants were encouraged to ask questions throughout the seminar. Finally, participants were allowed to interact with the tool and complete a TAM questionnaire.

\paragraph{Instrument} The study evaluates  two internal variables of TAM: perceived usefulness and perceived ease of use. Rationales  follow. As consultants of \textit{pure::variants}, the participants were in a good position to provide an informed opinion on how much they believe using \textit{FEACKER} would improve their job performance (perceived usefulness). As developers of \textit{pure::variants}, the participants could provide an informed opinion on the degree of ease in using \textit{FEACKER} for existing  \textit{pure::variants} users (perceived ease of use).

\paragraph{Results}  
Fig. \ref{fig:usefulness} displays the  chart for the variable `perceived usefulness'. Notice that `usefulness' is measured regarding the support given to the IF4SPLE process. The chart reveals a prevailing agreement among participants regarding the usefulness of \textit{FEACKER}. However, the chart also indicates  there  is no full consensus on this matter (this will later serve to inform the Focus Group). As for `perceived ease of use' (see Fig. \ref{fig:ease-of-use}), results tend to be more conclusive on the seamlessness to which \textit{FEACKER} is embedded within \textit{pure::variants}. The results suggest that using \textit{FEACKER} does not interfere with the existing GUI gestures  of \textit{pure::variants}.

\paragraph{Threats to validity}
\begin{itemize}

    \item \textbf{Construct Validity} refers to the level of precision by which the variables specified in research assess the underlying constructs of concern. Here, the constructs are `ease of use' and 'usefulness'. To mitigate its potential influence, the researchers utilized the TAM questionnaire.  To ensure internal consistency, Cronbach's alpha is calculated for the questionnaire, which resulted in an  $\alpha$ value of 0.62 and 0.97 for usefulness and ease of use, respectively. Cronbach alpha values of 0.7 or higher indicate acceptable internal consistency, with values above 0.9  perceived as redundancy of some questions. 

    \item \textbf{Internal Validity} refers to how much the independent variable(s) or intervention was the actual cause of the changes observed in the dependent variable. Here, the intervention is \textit{FEACKER}. Yet, other factors besides \textit{FEACKER} might influence the results. First, the background of the participants. In this regard, the researchers consulted employees of \textit{pure-systems Gmbh}. Except for one, all participants have at least two-year experience in both SPL and \textit{pure::variants} (see demographics at Fig. \ref{fig:demographics}). Second, the clarity and understandability of the questionnaire. To mitigate its potential influence, we resort to the TAM questionnaire, which has been widely employed and validated in assessing software engineering artefacts \citep{Wallace2014}. 

    \item \textbf{External Validity} assesses to what extent the results obtained in this study can be generalized to other scopes different to the one approached in the study. We defer the discussion of this issue to Section \ref{sec:generalization}.
\end{itemize}

\subsection{Focus Group evaluation }
TAM measures \textit{intention to use}, but it is not based on \textit{the real use}, hence, contradicting the second principle of an authentic evaluation \citep{Sjoberg02}. To mitigate this drawback, we resort to a second evaluation, now based on Focus Groups. 

\paragraph{Goal} The purpose of this study is to \textit{delve into} the divergent items of a previous TAM evaluation  with respect to  \textit{introducing IF4SPLE practices into pure::variants} from the point of view of \textit{annotated-based SPL practitioners}  in the context of the WACline SPL.

\paragraph{Participants} The group was formed by three engineers with at least two-year experience in using \textit{pure::variants}\footnote{Focus groups are typically made up of 3-12 people \citep{Parker2006}}. All of them were part of WACline's core-development team.

\paragraph{Design} Participants were asked to create a feedback model collaboratively. The model should approach analysis opportunities they were interested in. They came up with 18 different hits related to 11 features\footnote{The resultant feedback model can be found in a branch of WACline's repository: \url{https://github.com/onekin/WacLine/tree/featureTracking/WACline}}. Feedback was gathered during a week about WACline products running in a sandbox framework. So collected feedback was presented through the usage-based attributed feature model shown in Fig. \ref{fig:feedbackpull}.  Once the usage feedback was collected, the focus groups started. 

\paragraph{Instrument} The focus group was structured along those issues that rose the most significant divergence in the TAM evaluation (see Fig. \ref{fig:usefulness} and Fig. \ref{fig:ease-of-use}). Each issue was turned into a question for the focus group, namely:
\begin{itemize}
    \item \textbf{Is the feedback model able to capture your feedback needs effectively?} This question accounts for the TAM's US1 statement:  \textit{Using \textit{FEACKER} would enable me to accomplish IF4SPL tasks more quickly}. Fig. \ref{fig:usefulness} depicts the divergences in the answers: disagree (1), neutral (3), agree (3), and strongly disagree (1). Among the tasks introduced in the IF4SPL, the more labour-intensive one is feedback specification since it requires knowledge about  the feature model, the SPL product portfolio, and the code of the feature(s) at hand. Hence, the focus question delves into the extent FEACKER's feedback model can directly capture the needs of the analysts.
    \item \textbf{How seamlessly were IF4SPLE's new tasks integrated with your SPLE practices?} This question accounts for the TAM's US2 statement: \textit{Using \textit{FEACKER} would improve my performance following IF4SPL}. Fig. \ref{fig:usefulness} depicts the divergences in the answers: disagree (1), neutral (3), and agree (4). IF4SPLE proposes the need for an analysis space and the consequent tasks in SPLE. The way that these tasks are integrated into the traditional workflow may have a direct influence on the performance of the team. Hence, the focus question seeks to explore the impact of IF4SPLE tasks on the performance of domain engineers.
    \item \textbf{How seamless was \textit{FEACKER} integrated  with \textit{pure::variants}' gestures?} This question accounts for the TAM's EU1 statement: \textit{ Learning to operate \textit{FEACKER} would be easy for me}. Fig. \ref{fig:usefulness} depicts the divergences in the answers: strongly disagree (1), neutral (2),  agree (3), and strongly agree (2). Being \textit{FEACKER} an extension can disrupt the way of interacting with \textit{pure::variants}. Thus, this question tries to determine how much \textit{FEACKER} disturbs existing \textit{pure::variants}' gestures.
\end{itemize}

\paragraph{Results}
\begin{itemize}

    \item \textbf{Is the feedback model able to capture your feedback needs effectively?} Participants indicated that most of their feedback expressions refer to a single feature. No case popped up with AND/OR expressions. Two participants missed the ability to record values of variables at execution time. This would allow capturing the number of items a user has at the same time to operate within the GUI or capture values that could make the system fail.
    \item \textbf{How seamlessly were IF4SPLE's new tasks integrated with your SPLE practices?}Participants agreed on  feedback functionality  being subject to the same reuse principles as domain functionality  and hence, being moved to DE. As one participant put it: ``If testing is mainly a DE activity, why should feedback analysis differ? After all, usage feedback informs feature evolution, sort of perfective maintenance''. This said, they expressed two worries. First, the feedback model is fragile  upon changes in the \textit{\#ifdef} blocks. This introduces a dependency between code developers and feedback analysts. How severe is this limitation? Participants observe that feedback analysis tends to be conducted for mature features where customer usage drives feature evolution rather than feature `upbringing'. For mature features, the expectation is that code would be relatively stable and hence, with a low risk of interference between feedback analysis and feature developers.  Second and most important, upgrades on the feedback model would only take effect by deriving  the products again. This might be a severe concern for some installations, but, after all, this is what continuous deployment is supposed to be. Finally, some exciting issues arose with no clear answer: How is the evolution of feedback models conducted? How would different feedback models co-exist? Should variability be engineered within the feedback model itself? What strategy would be to face large feature models with intensive feedback needs? Should techniques similar to those for SPL testing be used?
    \item \textbf{How seamless was \textit{FEACKER} integrated  with \textit{pure::variants}' gestures?} Participants mostly appreciate conducting feedback analysis without leaving \textit{pure::variants}. This seems to suggest that variability managers should not overlook this concern. 
\end{itemize}

Nevertheless, some issues emerged:
\begin{itemize}
    \item Feedback Model. All participants found it cumbersome to specify. Difficulties came from Feedback Models' dependency on  the \textit{\#ifdef} blocks, specifically, the anchor specification. This limitation might be eventually mitigated through dedicated editors,
    \item Feedback Transformation. Participants found the approach intuitive and akin to the \textit{modus operandi} of \textit{pure::variants},
    \item Attributed Feature Models. Participants foresaw the benefits of introducing feature usage for more sophisticated feature analysis. However, feature models are very poor as  dashboards. Participants indicate the need to enhance variability managers with this functionality\footnote{It rests to be seen whether \textit{pure::variants}' tabular view of features which includes attributes, might provide a better fit.} or, instead, tap into existing dashboard applications through dedicated connectors.
\end{itemize}
\subsection{Threats to validity}
We follow here the threats to validity for Qualitative Research as described by  \citet{maxwell1992}
\begin{itemize}

    \item \textbf{Descriptive Validity} refers to the accuracy and completeness of the data. To mitigate this threat, we took notes about participants’ reactions as well as recording the session's audio to facilitate further analysis. Moreover, we asked for clarifications from the participants during the discussion. Another limitation is the small number of participants. This is partially due to the pursuit of realism in selecting the participants, who were required to be knowledgeable about both \textit{pure::variants} and GA. Favouring realism versus quantity is supported by the literature on focus groups, which recommends purposive sampling as the method for participant recruiting \citep{Morgan1997-vm}.
    \item \textbf{Interpretive Validity} refers to how well researchers capture participants’ meaning of events or behaviours without imposing their  perspective. It should be noted that the problem was pointed out by the practitioners themselves when attempting to accord \textit{pure::variants} and GA. To improve interpretative validity, we began the session with a brief presentation on the study's objectives and the proposed intervention. This presentation was aimed at establishing common terminology and avoiding misunderstandings.
    \item \textbf{Reproducibility}. \textit{FEACKER} and WACline are readily accessible to the public through download. The additional infrastructure utilized in this study, including \textit{pure::variants} and GA, is widely used among practitioners, facilitating the dissemination of this work's results.
\end{itemize}

\section{Generalization}\label{sec:generalization}

Following DSR, the situated learning from the research  project should be further developed into general solution concepts for a class of field problems \citep{Sein2011}. 
Sein et al. suggest three levels for this conceptual move:
\begin{itemize}
\item generalization of the problem instance, i.e., to what extent is \textit{implicit feedback} a \textit{problem} for SPL other than WACline?
\item generalization of the solution instance, i.e., to what extent is \textit{FEACKER} a \textit{solution} to implicit feedback in SPLs?
\item derivation of design principles, i.e., what sort of design knowledge can be distilled from the \textit{FEACKER} experience that might inform variability managers other than \textit{pure::variants}.
\end{itemize}
The rest of this section tackles these questions.

\subsection{Generalization of the Problem Instance} 
\newcolumntype{Y}{>{\centering\arraybackslash}X}

\begin{table*}[th!]
\scriptsize
\caption{Contextual characterization of the local experience in WACline. \label{tab:Contextual-data-along}}

\begin{tabularx}{\textwidth}{|l|Y|Y|}
\hline
  \multirow{4}{*}{Technical environment} & Programming language  & \textit{JavaScript}                    \\ \cline{2-3} 
                                                                & Branching strategy        & Grow-and-prune                         \\ \cline{2-3} 
                                                                & Variability manager       & \textit{pure::variants}                         \\ \cline{2-3}
                                                                & Tracker Manager       & \textit{Google Analytics}                         \\ \cline{1-3}
                                \multirow{4}{*}{SPL Attributes} & Lifespan                  & 5 years                             \\ \cline{2-3} 
                                                                & Size (approx.)             & 85 features \& 7 products             \\ \cline{2-3} 
                                                                & Domain                    & Web                                              \\ \cline{2-3} 
                                                                & Variability model        & Annotation-based                            \\ \cline{1-3}
                                \multirow{1}{*}{Implicit Feedback} & Purpose                  &  Scoping                             \\     
                                                                 \hline
\end{tabularx}

\end{table*}
Generalizing from a local experience starts by identifying the contextual parameters that might have influenced the intervention and its utility. Table \ref{tab:Contextual-data-along} gathers what we believe are the main contextual parameters that frame our case study. The question arises about whether our setting is somehow unique or rather other SPL installations can share it. 

The first consideration is the domain: the Web. Certainly, the Web is a pioneering domain in applying continuous deployment  using implicit feedback: mobile-app development  \citep{Xiao2021}; online training platform \citep{Astegher2021};  or
mobile and web contexts \citep{Oriol2018}.  However, we argue that the interest in implicit feedback is not limited to the Web. Clements's popular definition of an SPL  as addressing `a particular market segment' entails that the evolution of `this market segment' should, naturally, go hand in hand with the evolution of the SPL \citep{Clements2002}. Hence, SPL scoping (i.e., deciding on the features to be covered by the SPL)  is necessarily `continuous' to keep pace with this market segment. If SPL scoping is continuous, then implicit feedback will become a main informant of how this market segment evolves.   And if so, chances are the problem of implicit feedback found in WACline is  shared by  other installations. 

Notice  that WACline's incentive for implicit feedback is in informing scoping.  However, usage data is also useful for troubleshooting, fault prediction, supporting network operations or development activities \citep{Dakkak2021b,Dakkak2022b}, and enabling continuous experimentation \citep{Mattos2018}.  Specifically,  collected data provides valuable insights into the real-world behaviour and performance of embedded systems, allowing for the identification of potential issues and the development of solutions to improve performance. Additionally, models and simulations used in the evolution of embedded systems can be validated and refined, providing a more robust and accurate understanding of the system behaviour \citep{Dakkak2021b,Dakkak2022b}. We can then conjecture that implicit feedback would also be of interest to cyber-physical systems.

\subsection{Generalization of the Solution Instance} 

\textit{FEACKER} is an intervention for \textit{pure::variants} as the event manager,  GA as the event tracker,  Web-based product portfolios, and integrated product deployment. The following paragraphs discuss the limitations of such characterization in the generalizability of \textit{FEACKER}.

\paragraph{Technological limitations} The focus on \textit{pure::variants} limits our solution to annotation-based SPLs. Other approaches using  feature-oriented programming or component-based SPLs would need to resort to other means as our approach is heavily based on pre-compilation directives. As for GA, \textit{FEACKER}'s Feedback Model  might be biased by hit specification in GA.  Other  trackers like \href{https://grafana.com/}{Grafana }, \href{https://matomo.org/}{Matomo} or \href{https://www.zoho.com/}{Zoho} might have different expressiveness, which would involve changes in the Feedback Model. 

\paragraph{Organization limitations} \textit{FEACKER} is conceived for organizations with  access to SPL variants once deployed.  Yet, to collect usage data, a data pipeline must be established that connects to the deployed variants within the customer's network. These variants might reside in a protected intranet with multiple security gates and checkpoints to prevent unauthorized access. The access setup for each customer will be unique and require customized configurations to connect the variants to the data collectors and into the pipeline. This, in turn, raises the issue of data completeness. Suppose access to variant usage is limited, and only a subset of variants can be sampled. In that case, data completeness refers to how well the collected data represents the full range of products used by customers. Collecting data from functioning nodes is necessary for accurate results in some situations. However, this may not be possible in SPLs with hundreds of deployed variants due to technical or resource constraints. The challenge in these cases is to gather data from a representative sample of variants that accurately reflects the SPL platform.

\paragraph{Domain limitations} \textit{FEACKER} is an intervention for Web-based product portfolios. Using \textit{FEACKER} in  embedded and cyber-physical domains  introduces additional challenges: 
\begin{itemize}
    \item \textit{Performance impact.}  The limited resources of embedded and cyber-physical systems make it crucial to optimize data collection, as these activities can consume internal resources. One  solution  is collecting data during low-traffic hours. The casuistic can be exacerbated in SPLs, where each variant configuration may suffer different performance impacts, calling for a bespoken solution to minimize the performance impact on each deployed variant \citep{Dakkak2022a,Mattos2018}.
    \item \textit{Data dependability.} Embedded systems produce various data types, such as sensor readings, network status, performance metrics, and behaviour patterns. Evaluating just one type of data does not give a full understanding of its quality. While specific quality measures, such as integrity, can be applied to individual data types, a complete understanding of data quality requires analyzing the interrelationships and correlations between multiple data sources. The complexity arises when each product configuration has different sensors and data sources, requiring the feedback collection process to be aware of the unique settings and requirements of each configuration \citep{Dakkak2021b}.

\end{itemize}



\subsection{Derivation of design principles}
\begin{table*}[ht!]
\scriptsize
 \caption{Design Principles for  Implicit Feedback in Variability Managers.}
 \centering
\begin{tabularx}{\textwidth}{|p{0.5cm}|X|X|X|}
\hline
\textbf{ID} & \textbf{Provide \textit{variability managers} with...} & \textbf{for Domain Engineers to...} & \textbf{Realization in \textit{pure::variants}} \\ \hline\hline
DP1 & ... a goal-centric feedback specification model & ... structure feedback directives using  GQM  & a  \textit{yaml} realization of GQM constructs\\ \hline
DP2  & ... a feature-based  feedback  model & ... align feedback with other SPLE tasks   &  yaml's \textit{target} clause in terms of features \textit{} \\ \hline
DP3 & ... feedback transformations & ... account for `separation of concerns' to avoid polluting the platform code & A two-step Product Derivation  \\ \hline
DP4 &  ... feature-based feedback dashboards & ... track   SPL-wide metrics & Feature models with derived attributes\\ 
\hline
\end{tabularx}
\label{tab:design-principles}
\end{table*}
Design principles reflect knowledge of both IT and human behavior  \citep{Gregor2020}. Accordingly, a design principle should provide cues about the effect (e.g.., allowing for implicit feedback at the platform level), the cause (e.g., feedback transformation), and the context where this cause can be expected to yield the effect for the target audience (e.g.,  domain engineers).  Table \ref{tab:design-principles} outlines the main design principles. We consider as main design decisions the following: characterizing the feedback model in terms of features; structuring the feature model along the GQM model; and supporting the feedback model as a full-fledged model and hence executable by transformation.  

\section{Conclusion}\label{conclusions}
Implicit feedback is widely recognized as a key enabler of continuous deployment for one-off products. This work examined how feedback practices can be extended to the level of the SPL platform, affecting both SPLE processes and  variability manager tools. We advocate for placing features at the centre of the analysis and realizing feature tracking as a transformation approach. We developed \textit{FEACKER} as a proof-of-concept, which is available for public inspection and use. The evaluation included both real practitioners in a TAM evaluation (n=8) and a focus group (n=3). The results suggest that the approach is seamless with respect to current practices but raises several issues regarding its generalizability.
 
This work sought a  global perspective in introducing feedback analysis in SPLs. Each of the feedback tasks (i.e.,  specification, transformation, gathering and analysis)   is a subject in its own right, raising diverse questions: 
\begin{itemize}
\item How could feature usage leverage existing feature-analysis techniques? Could dependency between two features be weighted based on the (co)usage of these two features? Could this co-usage between features be used by configuration assistants to recommend co-features?
\item How could SPL evolution become `continuous' by tracking feature usage? Which sort of metrics can be most helpful in tracking SPL evolution? 
\item How can existing experiences on analysis dashboards be tuned for platform-based feedback? 
\item To what extent does implicit feedback account for the same benefits as in  one-off development?
\end{itemize}
Ultimately, we hope  to promote debate among  researchers and practitioners about the challenges of introducing implicit feedback in variability-rich systems. At play, bringing continuous deployment into SPLE.

\section*{Acknowledgement}
We would like to extend our gratitude to both WACline's and \textit{pure-systems}' employees for their invaluable contributions to this research. We are grateful for their time and effort. This work is co-supported by MCIN/AEI/10.13039/501100011033 and the “European Union NextGenerationEU/PRTR” under contract
PID2021-125438OB-I00. Raul Medeiros enjoys a doctoral grant from the Ministry of Science and Innovation - PRE2019-087324. 

\appendix
\section{Algorithms}\label{ap:algorithms}
\textit{Feedback Transformation (Algorithm \ref{alg:feedback-transformation})}: The process  starts by creating a copy of the platform code. Then, it filters the feedback directives specified in the feedback model (\textit{yaml} file) based on the expressions that match the configuration model. The corresponding event specifications are processed for the directives that pass this filter. The event pointcuts are determined using the file names and anchors defined in the feedback model, with the file name corresponding to an artifact of the Family Model and the anchor serving to single out the code fragment where the GA hit is to be injected. Finally, \textit{FEACKER} generates the GA hits and inserts them into the appropriate code lines.

\textit{Feedback Analysis (Algorithm \ref{alg:feedback-analysis}):} The algorithm gets as input a feature model. First, the analysis component searches for GA query parameters in the given feature model. The value of these attributes (i.e., the GA API query) is then used as the input to call the GA API. Finally, a copy of the feature model is created and the raw results obtained from GA API replace the query parameter value of the attributes in the cloned feature model.
\begin{algorithm*}[b!]

\caption{Feedback Transformation takes a feedback model, the platform code and a variant configuration model as an input and returns the platform code with embedded feedback hits.\label{alg:feedback-transformation}}
{\begin{lstlisting}[language=Java]
function PlatformCode feedbackTransformation( FeedbackModel feedbackModel,  PlatformCode platformCode, VariantConfigurationModel variantConfigurationModel){
    //Extract selected and deselected features from the variant configuration
    Map<Feature,Boolean> selectedFeatures = extractFeatures(variantConfigurationModel)
    //For each goal check whether the input variant is part of the goal context and if it satisfies the target expression
    foreach(goal in feedbackModel.goals){
        if(isExpressionTrue(feedbackModel.context, selectedFeatures) and isExpressionTrue(feedbackModel.target, selectedFeatures)){
            //Create the directory for the platform code with the embedded GA code
            Platform feedbackPlatformCode = copyDirectory(platformCode)
                //For each question in the goal, get its metrics 
                foreach (question in goal.questions){
                    foreach (metric in question.metrics){
                        //For each pointcut  find where the GA code should be injected
                        foreach (pointcut in metric.pointCuts){
                            File file = findFile(feedbackPlatformCode, pointcut.fileName)
                            foreach (ifdefBlock in file){
                                //Check if the #ifdef block satisfies the target
                                if(doesIfdefMatchExpression(feedbackModel.target, ifdefBlock)){
                                        //Generate the google analytics hit code and add it to the given file
                                        Integer anchorLine = findAnchor(pointcut.anchor, ifdefBlock, file)
                                        addGoogleAnalyticsHit(anchorLine, metric, file)
                                }
                            }
                        }
                    }
                }
            }
        }   
    }
    return feedbackPlatformCode
}

\end{lstlisting}}{\small\par}

\end{algorithm*}
\clearpage

\begin{algorithm*}[t!]

\caption{ Feedback Analysis, takes a feature model with GA query parameters on it and returns a cloned feature model with the raw results of those GA query. \label{alg:feedback-analysis}}
{\begin{lstlisting}[language=Java]
function FeatureModel feedbackPull(FeatureModel implicitFeatureModel){
    //Search for Google Analytics usage attributes in the implicit feature model.
    Map<String, String> usageAttributeMap = findFeedbackQueryParameters(implicitFeatureModel)
    Map<String, String> feedackResultsAttributeMap = new Map()
    //For each attribute, get its value, call GA API and store the result
    for each (usageAttribute in usageAttributeMap){
        String googleAnalyticsQuery = usageAttribute.value
        String feedbackData = googleAnalyticsApi.getFeedbackData(googleAnalyticsQuery)
        //Store GA API results in an auxiliary map
        feedackResultsAttributeMap[usageAttribute.name]=feedbackData
    }
    //Create the cloned feature model
    FeatureModel clonedFeatureModel=copyFeatureModel(implicitFeatureModel)
    //For each attribute replace the value with the result obtained from GA API
    for each (feedbackResultAttribute in feedackResultsAttributeMap){
        featureModelAttribute = findFeedbackAttributeInFeatureModel(feedbackResultAttribute.name, usageAttributedFeatureModel)
        featureModelAttribute.setValue(feedbackResultAttribute.value)
    }
    //Save the changes of the cloned feature model
    usageAttributedFeatureModel.saveChanges()
    return clonedFeatureModel
}
\end{lstlisting}}{\small\par}

\end{algorithm*}

\bibliographystyle{elsarticle-harv} 
\bibliography{splcSCOPIOUS}

\end{document}